\documentclass[twocolumn,tighten,times]{aastex62}
\usepackage{amsmath}
\usepackage{amssymb}
\usepackage{graphicx}				
\shorttitle{Frequency Dependence of Scintillation Arc Thickness.}
\shortauthors{Stinebring, D.\ R. , Rickett, B.\ J., \& Ocker, S.\ K.}

\def\simless{\mathbin{\lower 3pt\hbox
   {$\rlap{\raise 5pt\hbox{$\char'074$}}\mathchar"7218$}}} 
\def\simgreat{\mathbin{\lower 3pt\hbox
   {$\rlap{\raise 5pt\hbox{$\char'076$}}\mathchar"7218$}}} 
\def\veffvec{{\bf V_{\rm eff}}}

\def\fD{f_{\rm D}}
\def\DfD{\Delta f_{\rm D}}
\def\dfD{\delta f_{\rm D}}

\def\Veff{V_{\rm eff}}
\def\vpsrvec{{\bf V_{\rm psr}}}
\def\Stwo{$S_2(f_{\rm D}, \tau)$}

\def\btheta{{ \mbox{\boldmath $\theta$} }}      
\def\Btheta{{$B$(\mbox{\boldmath$\theta$}) }}

\def\be{\begin{equation}}
\def\ee{\end{equation}}
\newcommand{\bea}{\begin{eqnarray}}
\newcommand{\eea}{\end{eqnarray}}

\begin{document}

\title{The Frequency Dependence of Scintillation Arc\\Thickness in Pulsar B1133+16}

\author{Dan R. Stinebring}
\affiliation{Department of Physics \& Astronomy, Oberlin College, Oberlin, OH 44074}

\author{Barney J.\ Rickett}
\affiliation{University of California, San Diego, CA 92093}

\author{Stella Koch Ocker}
\affiliation{Department of Physics \& Astronomy, Oberlin College, Oberlin, OH 44074}
\affiliation{Department of Astronomy, Cornell University, Ithaca, NY 14853, USA}

\email{dan.stinebring@oberlin.edu}

\begin{abstract}
Scintillation arcs have become a powerful tool for exploring scattering in the ionized interstellar medium.
There is accumulating evidence that the scattering from many pulsars is extremely anisotropic resulting in highly elongated, linear brightness functions.
We present a three-frequency (327~MHz, 432~MHz, 1450~MHz) Arecibo study of scintillation arcs from one nearby, bright, high-velocity pulsar, PSR~B1133+16.
We show that a one-dimensional (1D), linear brightness function is in good agreement with the data at all three observing frequencies.
We use two methods to explore the broadening of the 1D brightness function \Btheta as a function of frequency:
1) crosscuts of the forward arc at constant delay and 2) a 1D modeling of \Btheta using a comparison between model and observed secondary spectrum as a goodness-of-fit metric.   {Both methods show that the half-power width of \Btheta deviates from the expected dependence $\propto \nu^{-a}$, where $\nu$ is the observing frequency .   Our estimates of $a$ have moderately large uncertainties but  imply $a \lesssim1.8$, and so are inconsistent with the expected $a = 2.0$ for plasma refraction or $a = 2.2$ for Kolmogorov turbulence.   In addition the shape of \Btheta cuts off more steeply than predicted for Kolmogorov turbulence.   
Ultimately, we conclude that the underlying physics of the broadening mechanism remains unexplained.}
Our results place the scattering screen at a distance that is broadly consistent with an origin at the boundary of the Local Bubble.

\end{abstract}

\keywords{pulsars: individual (B1133+16) --- ISM: individual objects (Local Bubble) --- local interstellar matter}

\section{Introduction}

\nocite{smc+01}
In the discovery paper (Stinebring et al.\ 2001, hereafter S+01), scintillation arcs
were explained through the interference of multiple rays deflected by small-angle scattering
in the ionized interstellar medium.
Subsequent radio observations of the parabolic arc phenomenon 
(Hill et al.\ 2003; Hill et al.\ 2005; Cordes et al.\ 2006, hereafter C+06; Stinebring 2006; Putney and Stinebring 2006; Stinebring 2007a; Stinebring 2007b; Trang and Rickett 2007; Hemberger and Stinebring 2008; Brisken et al.\ 2010; Pen et al.\ 2014; Bhat et al.\ 2016; Liu et al.\ 2016)
have revealed a distribution of compact plasma structures in the interstellar medium on scales of 1-10 A.U., which also contain much finer structure that scatters the radio waves from pulsars into remarkably elongated (i.e. anisotropic) brightness distributions.  
However, 17 years later we still do not know the astrophysical nature of the structures responsible for the scattering.

Since the refractive index of the plasma depends strongly on frequency, how the arcs vary with the observing frequency provides important clues as to their origin.  Note however, that Hill et al (2003) \nocite{hsb+03} convincingly showed that the curvature of the arcs scales as the inverse square of the observing frequency, in agreement with the basic theory developed by S+01, which does not assume a plasma scattering mechanism.   The arcs are most clearly defined for relatively nearby pulsars and have been extensively observed in two of the earliest known pulsars, B0834+06 and B1133+16.  
Hill et al (2005) \nocite{hsa+05}
observed even finer structures in the ``reverse arclets'' of B0834+06, whose apex positions were independent of frequency, which requires highly localized regions of scattering that do not shift in sky position as a function of observing frequency.    

The primary observable quantity for these studies is the dynamic spectrum of the interstellar scintillation of a pulsar, which is a temporal sequence of high-frequency-resolution spectra of the pulsar, which is treated like a continuum source  \nocite{crsc06} averaged over brief subintegration times, typically 10 sec.  Arcs are revealed in the ``secondary spectrum'' \Stwo, which is the two-dimensional power spectrum, computed from the dynamic spectrum  with differential Doppler frequency, also called fringe frequency, ($\fD)$ and differential delay ($\tau$) as the variables conjugate to frequency and time, respectively.  Arcs are defined as distributions in delay and Doppler frequency that follow a parabolic form in \Stwo.   Often there is a ridge in $S_2$ along a simple forward parabola $\tau=\eta\fD^2$, where $\eta$ is the curvature.  The detailed shapes vary from pulsar to pulsar and with the observing frequency.
Some pulsar/frequency combinations also exhibit pronounced variability of the power distribution of $S_2$, particularly in cases where reverse arclets are present (Hill et al.\ 2003; Hill et al.\ 2005; C+06; Stinebring 2007b; Hemberger and Stinebring 2008).

The theory of scintillation arcs was elaborated by Walker et al. (2004) \nocite{wmsz04} and C+06.  
Scattering can be expressed as an angular spectrum of plane waves or scattered brightness distribution \Btheta, and scintillation is due to the mutual interference between all pairs of such wave components plus relative transverse motion between source and observer.  
Interference is only possible since all scattered waves are mutually coherent originating from an extraordinarily compact source.  
They show that in the special case where $B(\btheta)$ is one-dimensional (1D), the relationship simplifies greatly as discussed below in \S\ref{sec:1D-theory}.  While scintillation arcs can be observed in some specialized cases from isotropic 2D brightness distributions, narrow (thin) arc structures imply that the brightness distribution is highly anisotropic (elongated).

%
There is substantial evidence that some scintillation arcs arise from essentially 1D features on the sky where the underlying deflecting structures are seen in projection.
The best-studied case is that from the remarkable very long baseline interferometry observations of Brisken et al. (2010) \nocite{bmg+10}
of the pulsar B0834+06, which is known to display a wide range of scintillation arc phenomena (Hill et al.\ 2003;  Hill et al.\ 2005\nocite{hsa+05}; C+06).\nocite{crsc06}
As shown clearly in Figure~5 of their paper, the points on the sky that give rise to the scintillation arc are stretched out over more than 20~mas with an unobservably thin width perpendicular to the main axis.
Despite this one-dimensional structure on the sky, the scintillation arc is a thick one composed of multiple reverse arclets.  By thickness of the main forward arc we are referring to the broadening of the power distribution normal to the $\tau=\eta\fD^2$ locus.   In this paper we analyze multi-frequency arc observations recorded in total intensity at Arecibo from one particular pulsar B1133+16.  We find evidence that the brightness is nearly 1D, and study how the arc thickness varies with observing frequency providing additional insight into the underlying deflection mechanism.


In their study of three pulsars (B0834+06, B1133+16, and B1929+10) Hill et al.\ (2003) showed conclusively that the arc curvature scales as $\eta \propto \nu^{-2}$, where $\nu$ is the central observing frequency.  These three pulsars exhibit very thin arcs, which allows their curvature to be measured precisely.
Their arcs were thin enough to test the parabolic shape carefully and verify that they followed an $\eta \propto \nu^{-2}$ scaling.  
The implication is that the scattering structure is highly elongated along a straight line. 

In \S\ref{sec:theory} we present enough theory to explain the thickness of the scintillation arcs. 
In \S\ref{sec:Obs} we present our observations and  in \S\ref{sec:Analysis} two methods of arc thickness analysis.
In \S\ref{sec:DiscConc} we conclude that the scattered brightness distribution from B1133+16 is essentially 1D and discuss how the observed frequency-scaling in the arc thickness constrains the physics of the scattering.

\section{Theory of scintillation arcs from a single screen}
\label{sec:theory}

We concentrate on the scattering from a single localized region (screen), which creates a 2D brightness distribution $B(\btheta)$.  Scintillation is due to mutual interference between all pairs of waves scattered from the screen, observed arriving at angles $\btheta_1,\btheta_2$.  C+06 express the secondary spectrum \Stwo as a double integral over $B(\btheta_1)B(\btheta_2)$ times $\delta$-functions that constrain the angles to a particular $\fD, \tau$ combination :
\bea
\fD &=&  \nu\,(\btheta_1 -\btheta_2) \cdot \veffvec/c,   \label{eq:Doppler} \\
\tau &=& \frac{D_{\rm eff}}{2c}(\theta_2^2 - \theta_1^2). \;\     \label{eq:Delay}
\eea 
Here, the effective distance is $D_{\rm eff} = D_{\rm psr}(1-s)/s$ as given  in equation (2) of C+06, where $s$ is dimensionless parameter $0 \le s \le 1$, which is the ratio of the screen-pulsar distance to the Earth-pulsar distance ($D_{\rm psr}$).  However, note that their notation $d_e$ differs from our $D_{\rm eff}$, which uses the same notation as in equation (9) of Brisken et al. (2010).
The effective velocity is $\veffvec = [(1-s) {\bf V}_{\rm psr} + s  {\bf V}_{\rm obs} - s {\bf V}_{\rm scr}]/s$, where only  velocities transverse to the line of sight are important, ${\bf V}_{\rm obs}$ and ${\bf V}_{\rm scr}$ are the observer and screen velocities, respectively (Cordes and Rickett 1998).
\nocite{cr98}
As appropriate for B1133+16, we assume the special case that the proper motion velocity of the pulsar is much greater than the velocities of both the Earth and the scattering screen; hence,
\be 
\veffvec \approx \left(\frac{1-s}{s}\right) \vpsrvec.
\ee

\subsection{Forward and Reverse Arclets} 

In most cases $B(\btheta)$ decreases away from its maximum near the origin $\btheta=0$, which leads to a forward arc due to interference of the relatively unscattered wave components near the origin with those that are more highly scattered.  This case can be studied by setting, say, $\btheta_2=0$ in Equations~(\ref{eq:Doppler}) \& (\ref{eq:Delay}), which gives the constraint:
\be
\tau \ge \frac{cD_{\rm eff}}{2\nu^2V_{\rm eff}^2}\fD^2 .
\ee
Thus with $\btheta_2=0$, \Stwo lies above a simple bounding parabola for positive $\tau$ and below an inverted parabola at 
 delays which arises by setting $\btheta_1=0$.     In the general case interference must be considered between all pairs of scattered waves;  the angular extent of the scattered image along the effective velocity vector determines the extent in $\fD$ of arc features in the secondary spectrum, as noted in the discovery paper  (S+01).   In addition it is the width of the brightness function near the origin, which can be referred to as the ``core,'' that determines the thickness of the forward arc.    

See C+06 for examples of the secondary spectrum computed from various shapes assumed for the brightness distribution.  They conclude that prominent parabolic arcs imply anisotropic brightness distributions extended more along the velocity vector than perpendicular to it.  Further, a very narrow arc can only be due to a very anisotropic brightness distribution.  This leads us to consider 1D brightness distributions.  However, it does not follow that a thick arc is incompatible with a 1D brightness, as we discuss in \S\ref{sec:Analysis}.  

Arcs are also seen in some pulsars as isolated inverted parabolas (reverse arclets), whose apexes often lie along the main forward parabola.  These are due to isolated peaks in brightness offset from the origin that interfere with the waves scatterd from the core.


\subsection{Secondary Spectrum for 1D Brightness Profile}
\label{sec:1D-theory}

We now focus on the one-dimensional  case where $B(\btheta)=B(\theta_{\parallel})$.  Here $\theta_{\parallel}$ is the angular offset along a line that is oriented at angle $\psi$ to the effective velocity vector.  The secondary spectrum is then given by Equation~(13) of C+06, which in our notation becomes:
\bea
S_2(\fD,\tau) &=& B(\theta_{+}) \, B(\theta_{-}) \; /|2\sqrt{\eta} \fD|, \label{eq:S21D} \\[10pt]
\mbox{where} \;\theta_{\pm} &=& \sqrt{\frac{c\eta}{2D_{\rm eff}}}[-\tau/(\eta \fD)\pm \fD] ,  \\
\eta &=& \frac{c D_{\rm eff}}{2\nu^2 V_\psi^2}  \;\;\; \mbox{and}  \;\; V_{\psi} = \Veff\cos\psi .
\label{eq:1Dtheory}
\eea 
Thus the secondary spectrum at any given point ($\fD,\tau$) is due to the interference between the particular pair of waves at angles $\theta_{\pm}$.  

In most cases $B(\theta_{\parallel})$ decreases away from its maximum near the origin $\theta_{\parallel}=0$, which leads to a forward arc from interference between power near the origin (the core) and the rest of the brightness distribution.
Setting either $\theta_{\pm}$ to zero defines the parabolic arcs $\tau =  \pm \eta \fD^2$; note that we normally concentrate on positive delays and ignore the secondary spectrum at negative delays, where $\fD$ is also reversed in sign.  The parabolas are thickened by the width of the peak in brightness around the origin.

Consider now a subsidiary maximum in brightness  at $\theta_p$. It forms a reverse arclet due to its interference with the rest of the brightness distribution:
\be  
\tau =  \eta [f_{\rm apx}^2 - (\fD - f_{\rm apx})^2] , \; \mbox{where } \; f_{\rm apx}= -\theta_{p} V_{\psi}\nu/c .
\ee
It follows that the apexes of all such arclets would lie on the forward parabolic arc $\tau = \eta\fD^2$.  
The secondary spectrum due to all of the interference terms originating at $\theta_p$ is:
\be
S_{2,p}(\tau,\fD) = B[(\fD- f_{\rm apx})c/(\nu V_{\psi})] \, B[\theta_{p}] \; /|2\sqrt{\eta} \fD|, 
\ee

Previous investigators (Brisken et al.\ 2010; Liu et al.\ 2016) have used this relation to analyze isolated reverse arclets and infer information about the overall brightness function.
\nocite{lpm+16}
The superposition of such reverse arclets from all scattered points like $\theta_p$ will appear as a thickened forward arc.  In the next section we define the arc thickness from ``crosscuts'' at constant delay, which allows us to estimate the width of the central peak in $B(\theta_{\parallel})$.   
  
\subsection{Crosscuts at constant delay}
In \S\ref{sec:cc} we analyze the arcs from crosscuts of $S_2$ as a function of $\fD$ at constant delay.  Here we reconsider Equation~(\ref{eq:S21D}) at fixed positive delay $\tau$ :
\be
S_{2,cc}(\tau,\fD) = B[\theta_{+}- \fD(\lambda/V_{\psi})] \, B(\theta_{+}) \;/|2\sqrt{\eta} \fD|  \;, 
\ee
where
\be
\theta_{+} = \sqrt{\frac{c\eta}{2D_{\rm eff}}}[-\tau/(\eta \fD) + \fD] 
\ee
One can see that with fixed $\tau$ and $\fD>0$, $\theta_{+}$ increases with increasing  $\fD$ and passes through zero at $\fD = \sqrt{\tau/\eta}$, where the crosscut passes the center of the positive parabolic arc.  At this point $B(\theta_{+})$ peaks, since we assume that the brightness is greatest near $\theta = 0$.  
In addition, $\theta_{-}$ reaches a maximum  at this point and is stationary versus $\fD$.   So we expand both $\theta_{\pm}$ to first order in the difference from the crossing point $\dfD = \fD -\sqrt{\tau/\eta}$ and obtain:
\be
S_{2,cc}(\tau,\fD)  \approx   B(\dfD\frac{c}{\nu V_\psi}) B(\sqrt{2c\tau/D_{\rm eff}})/2\sqrt{\tau} . 
\ee
Here the weak $1/\sqrt{\eta}\fD$ factor is approximated as $1/\sqrt{\tau}$.  The result shows that the profile of a constant-$\tau$ crosscut maps out the peak of the brightness function $B(\theta_{\parallel})$ according to the simple scaling:
\be
\theta_{\parallel} = \dfD\frac{c}{\nu V_\psi} = \dfD \sqrt{\frac{2\eta c}{D_{\rm eff}}}, 
\label{eq:DeltafD}
\ee
as expected from Equation~(\ref{eq:Doppler}).  
Note that in the second form $\theta_{\parallel}$ can be converted into milliarcseconds with no reference to the unknown angle $\psi$, 
since $\eta$ is a measured quantity.
In \S\ref{sec:cc} and \S\ref{sec:1dfits} we use a scaled version of this angle, $\theta_{\rm m}$, that differs from that in Eqn~\ref{eq:DeltafD} by using the known value $D_{\rm psr}$  rather than $D_{\rm eff}$, hence 
\be
\theta_{\rm m} = \theta_{\parallel} \sqrt{\frac{1-s}{s}}  .
\label{eq:theta_m}
\ee

\section{Observations}
\label{sec:Obs}
In the first half of 2015 we observed pulsar B1133+16 with the Arecibo Observatory (project P2952), maintaining an approximately weekly cadence for 21 epochs.
At each observing epoch we obtained about 45~min of data at 432~MHz, our primary frequency, and then, in alternating sessions, a similar observation at either 327~MHz or 1450~MHz.
In order of increasing frequency (327, 430, 1450), the bandwidths employed were 50~MHz, 25~MHz, and 160~MHz, respectively; in the same order, the number of spectral channels used was 4096, 2048, and 2048.
This resulted in a Nyquist frequency along the delay axis of 38.4~$\mu$s at the two lower frequencies and 6.4~$\mu$s at 1450~MHz.
Using the Mock\footnote{https://www.naic.edu/ao/scientist-user-portal/astronomy/backends} spectrometers, we sampled the spectra at an interval of $\Delta t \approx 4$~ms and stored the data on disk for offline processing.
Following the procedure outlined in Hill et al. (2003), we then formed ${\tt ON}(\nu)$ and ${\tt OFF}(\nu)$ spectra with respect to the pulse time of arrival and produced a final spectrum, accumulated for a subintegration of 10~s, of $S_1(t_i, \nu) = [{\tt ON}_i(\nu) - {\tt OFF}_i(\nu)]/\langle {\tt OFF}(\nu) \rangle$, where $i$ labels the subintegration number and $\langle {\tt OFF}(\nu) \rangle$, is an average over the entire observation that was used to correct for bandpass rolloff.

At the end of the program we had obtained useful data at 11, 18, and 8 epochs for 327, 432, and 1450~MHz, respectively.
Here we report on {5 data sets obtained over the MJD interval 57173 -- 57185: at 327~MHz we observed on MJD~57173 and 57185; at 432~MHz on MJD~57173 and 57179; and at 1450~MHz we observed on MJD~57179. We chose this interval because the scintillation arc structure was simple at all three frequencies, consisting of a single arc with a minimum of reverse arclets.}
We will be presenting the results of the full observing program elsewhere.

\begin{figure*}[htb]
\begin{center}
\begin{tabular}{lll}
\includegraphics[trim = 5 0 40 40, clip,angle=0,width=5.6cm]{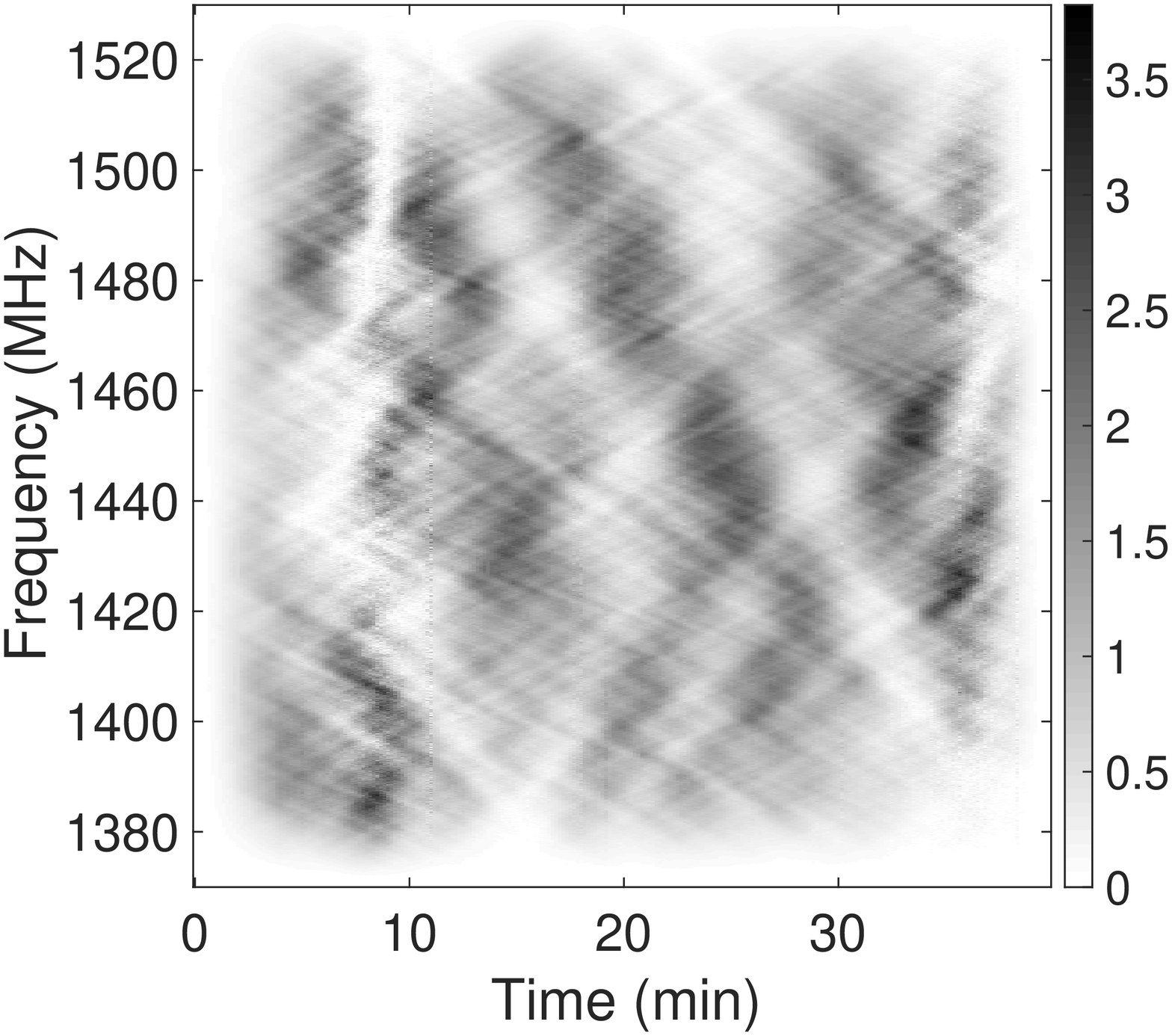} &
\includegraphics[trim = 5 0 65 40, clip,angle=0,width=5.4cm]{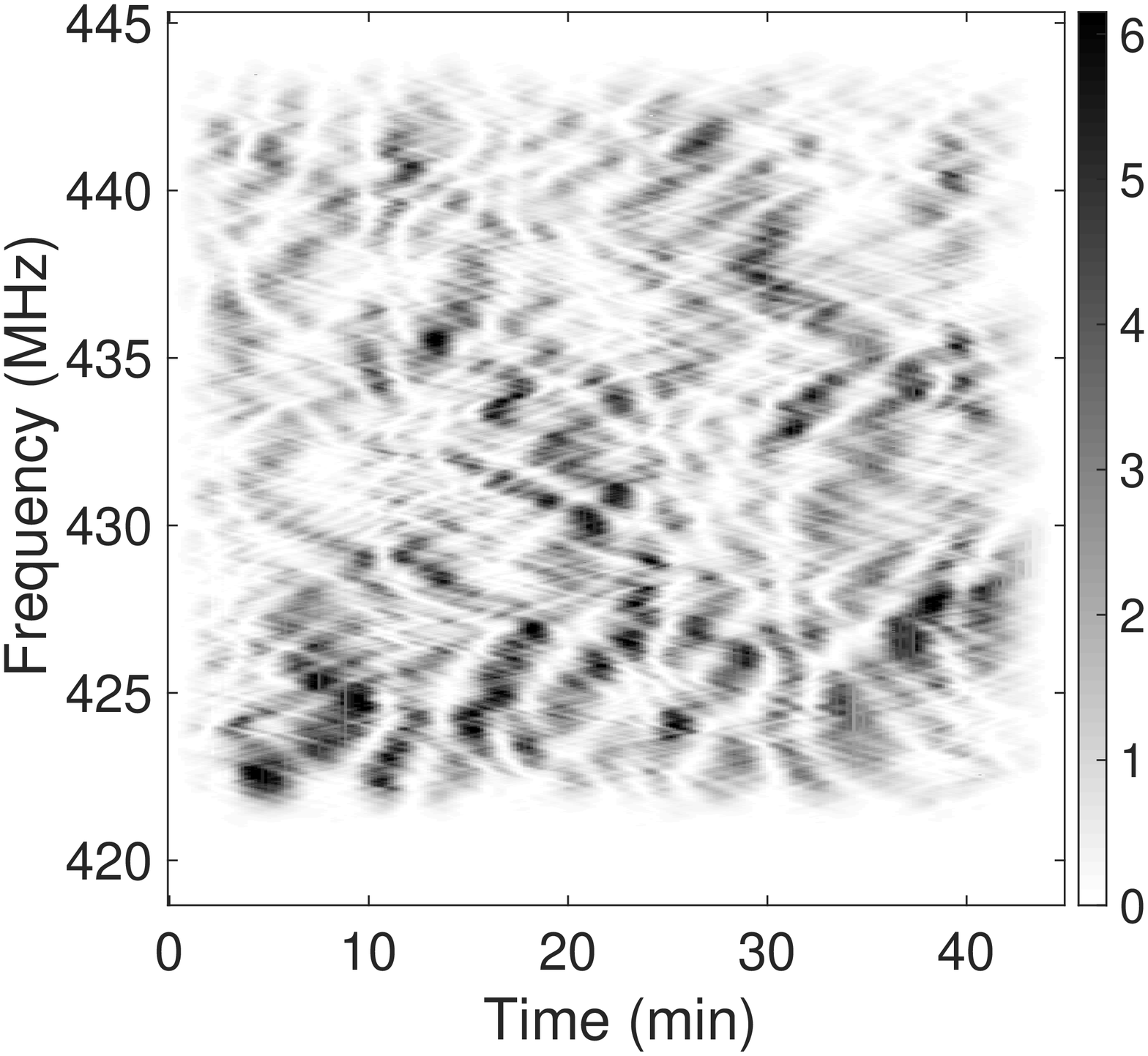} &
\includegraphics[trim = 5 0 65 40, clip,angle=0,width=5.4cm]{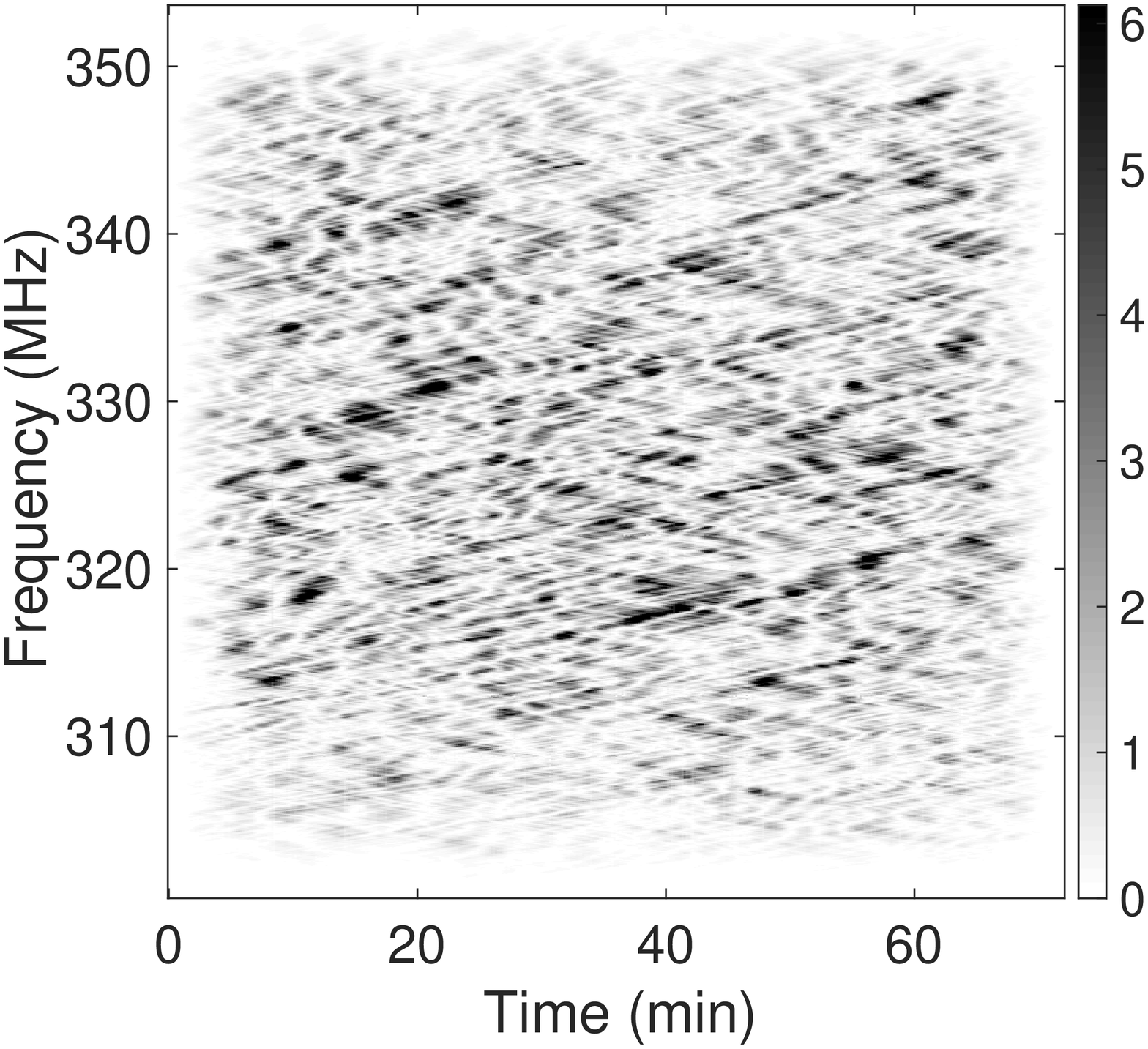} \\
\includegraphics[trim = 15 0 40 35, clip, angle=0,width=5.5cm]{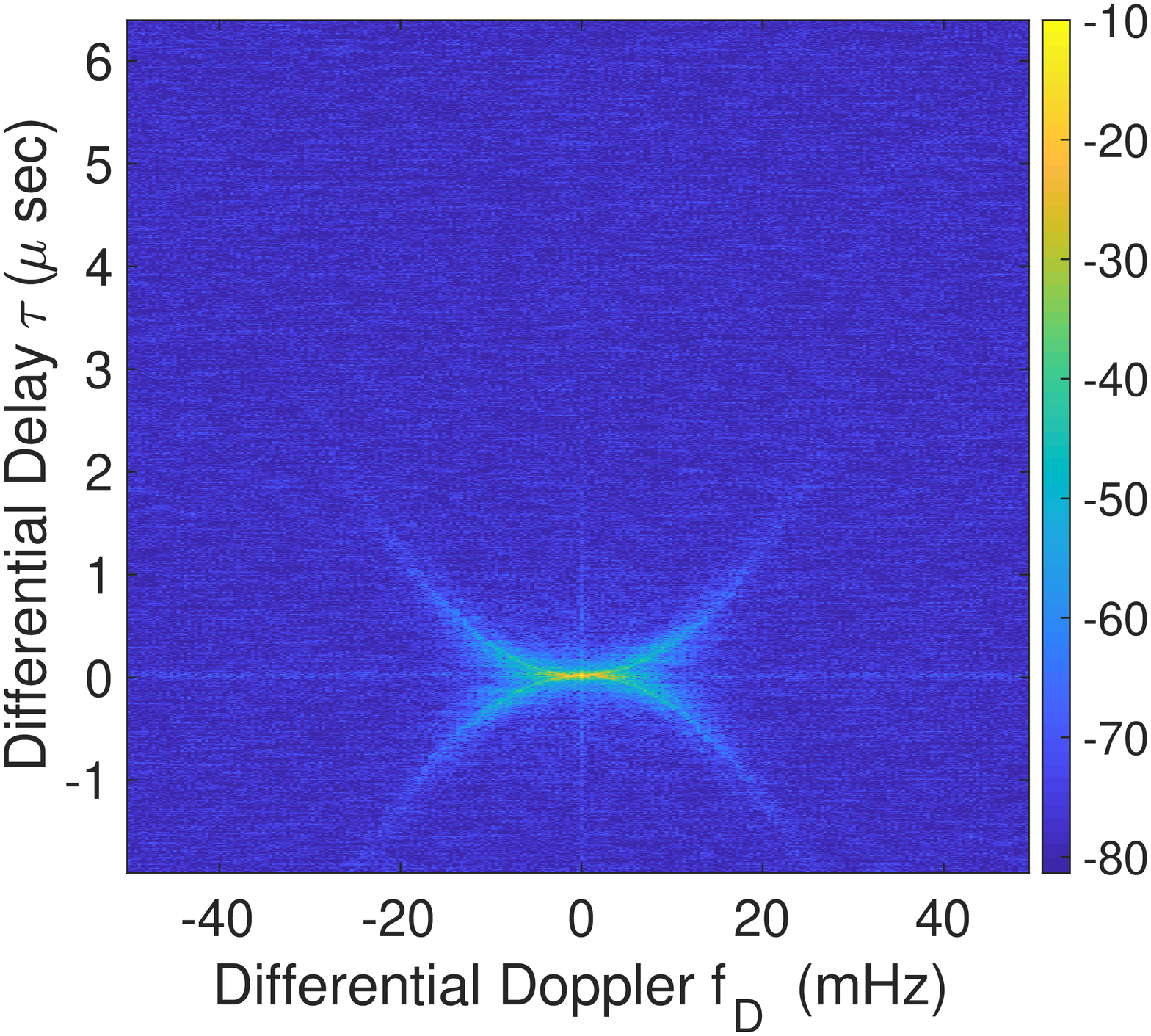} &
\includegraphics[trim = 5 0 40 35, clip, angle=0,width=5.5cm]{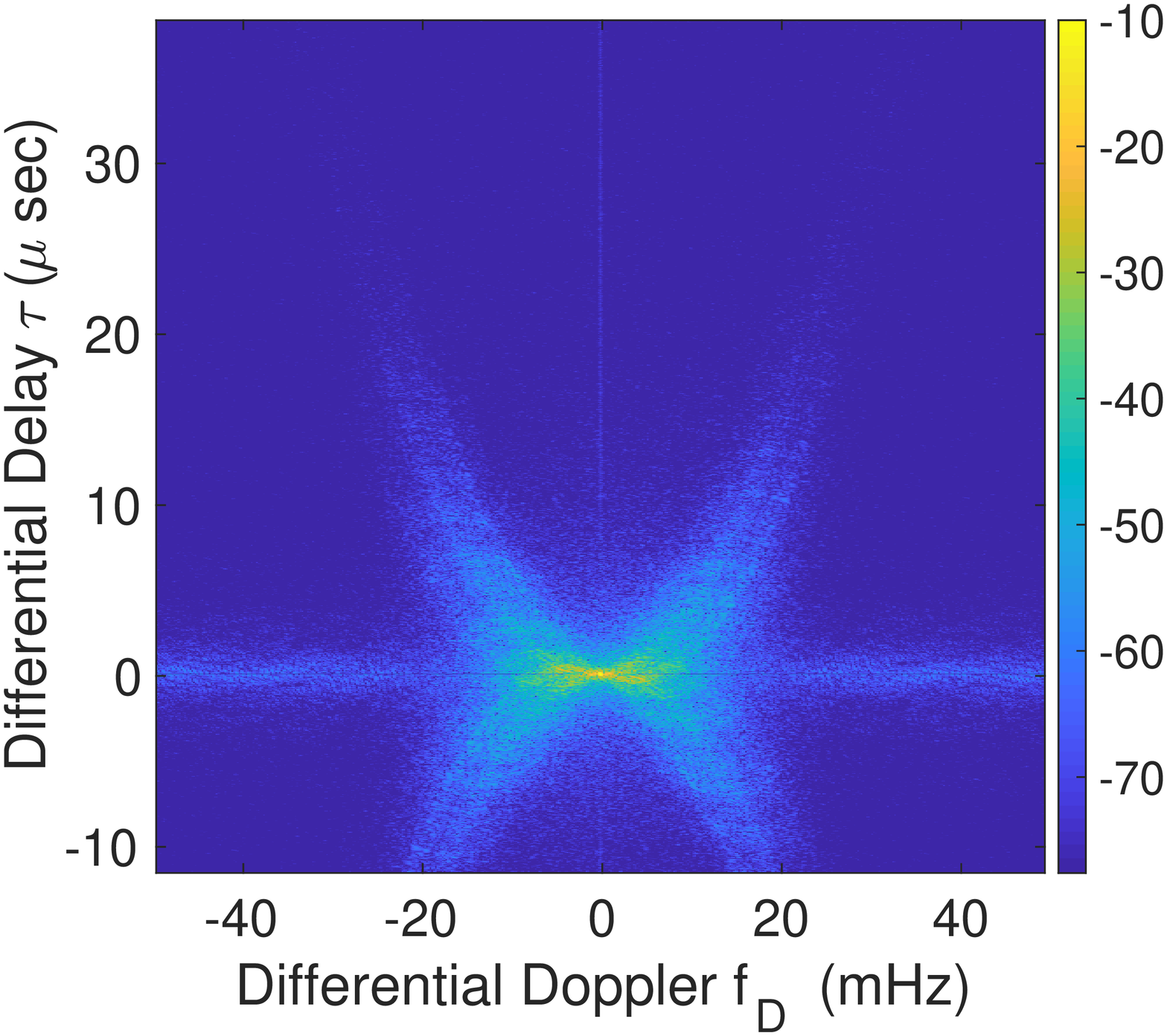}  &
\includegraphics[trim = 5 0 40 35, clip, angle=0,width=5.5cm]{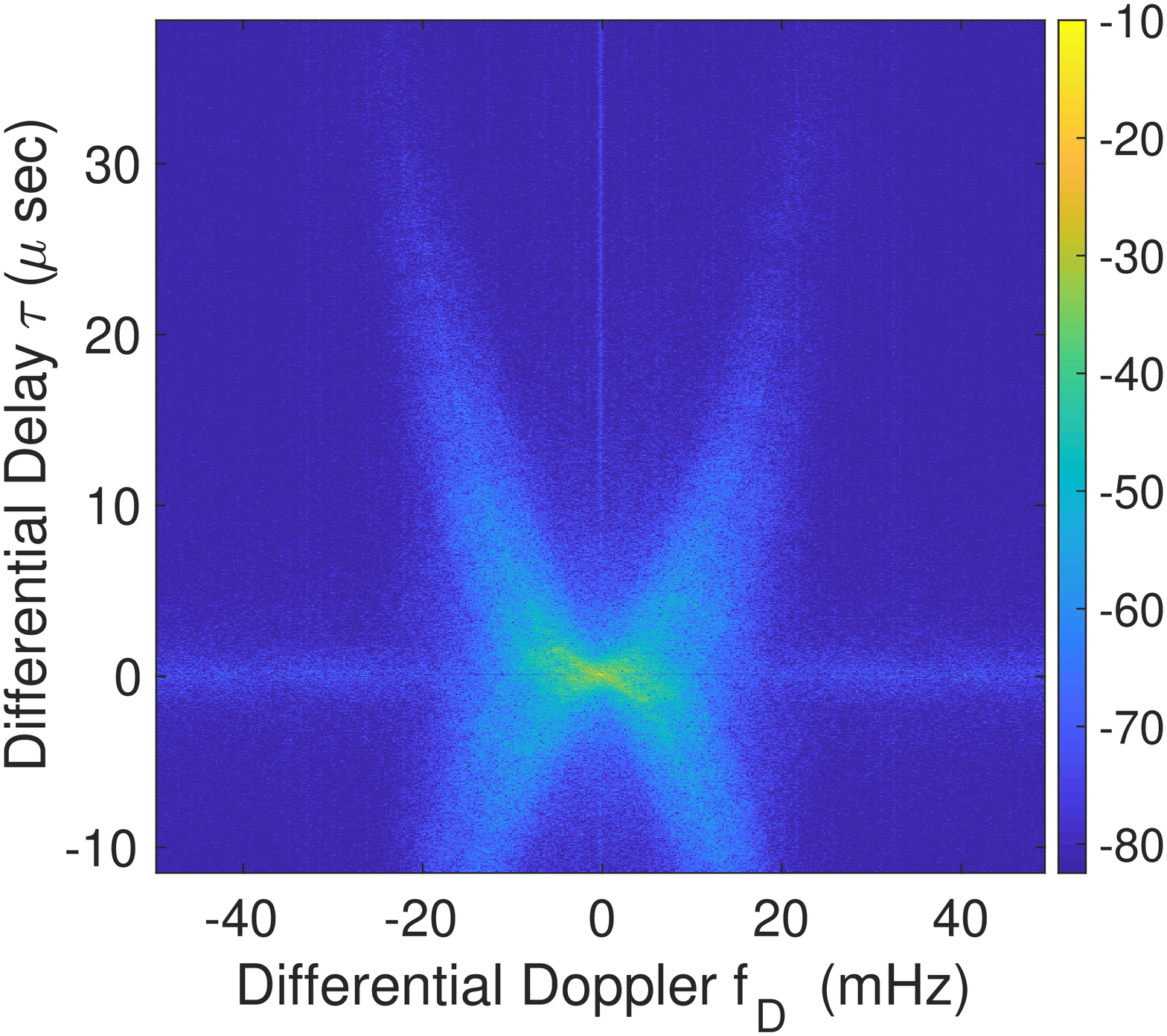}
\end{tabular}
\end{center}
\caption{Interstellar scintillation observed at the Arecibo Observatory from pulsar B1133+16.  Left: MJD 57179  at 1450 MHz. Middle:  MJD 57179  at 432 MHz. Right:  MJD 57173  at 327 MHz.  
Upper row:  Dynamic spectrum normalized to the overall mean.  The grayscale is linear in units of the overall mean.  A ``cosine-squared'' window has been applied to the outer 10\% of the spans in time and frequency, in order to lower the sidelobes in the spectral resolution function.
Lower row: Secondary spectra.  Colorbar scales are logarithmic (dB).   The low level stripe along the Doppler axis is due to imperfect correction of the broad band intrinsic variations.  The plot includes part of the negative delays in order to display the peak of the arcs more clearly.  Particularly at 432 MHz, this also shows how the forward arc is composed of reverse arclets.}
\label{fig:ds} 
\end{figure*}

In Figure~\ref{fig:ds} we show the dynamic and secondary spectra at the three frequencies for two closely spaced days, MJD~57173 (2015 May 31) and MJD~57179.
The upper row of panels shows the dynamic spectrum with a grayscale that is linear in power.  
These data have been normalized at each subintegration by the average pulse amplitude over the full bandwidth.  
This suppresses the broad-band intrinsic pulsar amplitude modulation which causes deep variations that are independent from one time step to the next.  
The dynamic spectra show deep modulation with narrow frequency structure, characteristic of interstellar scintillation (ISS).  There are also notable criss-cross patterns, whose character is revealed in the secondary spectra.  
From the dynamic spectrum at all 3 bands we formed the secondary spectra \Stwo, which we plot using a logarithmic grayscale (expressed in dB relative to the maximum power) in the lower panels of Figure~\ref{fig:ds}.  
The secondary spectra reveal well-formed forward scintillation arcs, which become thicker with decreasing frequency.
The details of this thickening form the main focus of the rest of the paper. 

Before turning to a more detailed analysis, we briefly discuss the chosen epoch with respect to the behavior of the scintillation arcs over the  21 epochs of the observing program.
This pulsar exhibits as many as four well-defined scintillation arcs (Putney and Stinebring 2006; Stinebring 2006), although a typical epoch usually shows between 1 -- 3 arcs.
\nocite{ps06a,sti06}
By analyzing archival data, we have shown that the curvature of these arcs has remained constant over time for more than 35 years (Stinebring et al.\ 2018, in preparation), reinforcing a model of multiple physical screens intercepted by the moving line of sight.
Since the arcs are much thinner at high frequency, the multiple arcs are best studied at 1~GHz and above for this pulsar.
At the outset of our program the pulsar was showing the three arcs labeled $b,c,d$ by Putney and Stinebring (2006).
This triplicity was most pronounced about one third of the way through our observations ($\sim$~MJD~57098), and the secondary spectra became dominated by the $b$ arc by the end of the program and in the results reported here.

Previously published secondary spectra for this pulsar (S+01; Hill et al.\ 2003; C+06; Trang and Rickett 2007) show occasional episodes in which one side of the scintillation arc is much brighter -- or extends much higher in delay -- than the other.
However, that was not the case for the five months of data collected in this program.
The secondary spectra shown in Figure~\ref{fig:ds} are typical of the entire data span in this regard: although {\em small} positive vs.\ negative $\fD$ asymmetries are present, particularly in terms of relative power levels, they are not a dominant feature in this data set.

\section{Analysis}
\label{sec:Analysis}
We have characterized the thickness of the forward arcs in two ways.  In the first approach (\S\ref{sec:cc}) we use crosscuts through the scintillation arc at fixed delay ($\tau$) followed by Gaussian fits to characterize the width of the arc.   In the second approach (\S\ref{sec:1dfits}) we fit a 1D brightness distribution model to the entire secondary spectrum. 
Before proceeding we determine the forward arc curvature and related quantities.  

The curvatures of the scintillation arcs were obtained using two techniques that gave consistent results. In one approach we used \texttt{parabfit} (Bhat et al.\ 2016), 
\nocite{bot+16}
based on the Hough transform used in image processing. 
We used another approach to determine uncertainties in $\eta$ at the two lower frequencies: we overlaid parabolas of given curvature on the secondary spectra and inspected them to determine the range of plausible $\eta$ values. 
The results of this combined analysis are shown in Table~\ref{tab:curvature}.
Using the values of $D_{\rm psr} = 357\pm 20$~pc and $V_{\rm psr} = 636\pm 40$~km~s$^{-1}$ (Brisken et al.\ 2002),\nocite{bbgt02} we tabulated values and uncertainties for $s_0$, the value of $s$ 
inferred when we assume $\psi = 0$.
(Note that the actual value will satisfy $s \leq s_0$; the screen will always be closer to the pulsar than for the case of $\psi=0$.)
The three $s_0$ values are consistent with each other, and we use their weighted mean of $s_0 = 0.62 \pm 0.01$ in what follows.
\begin{table}[htb]
\centering
\begin{tabular}{c  c    c}
freq (GHz)&	$\eta$ (s$^3$) &	$s_0$\\
\hline
0.327 &	0.06(1)&			0.61(4)\\
0.432&	0.040(5)&			0.65(3)\\
1.450&	0.0031(2)&			0.62(2)
\end{tabular}
\caption{Measurement of the arc curvature value $\eta$ from the data in Figure~\ref{fig:ds}. 
\label{tab:curvature}}
\end{table}

\subsection{Crosscuts at Constant Delay}
\label{sec:cc}
\begin{figure*}[htb]
\centering
\makebox[\textwidth][c]{
\includegraphics[width=0.33\textwidth]{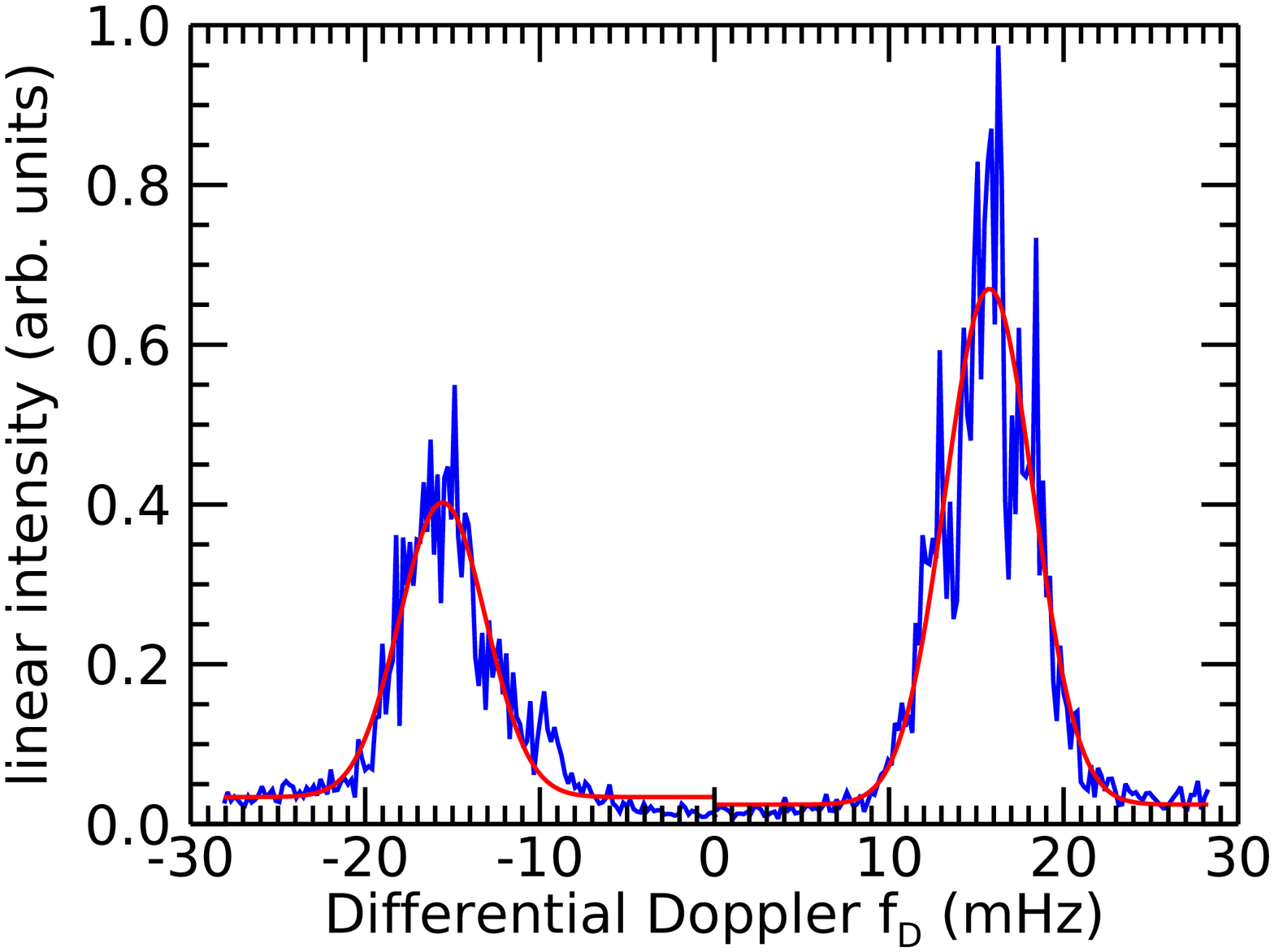} 
\includegraphics[width=0.33\textwidth]{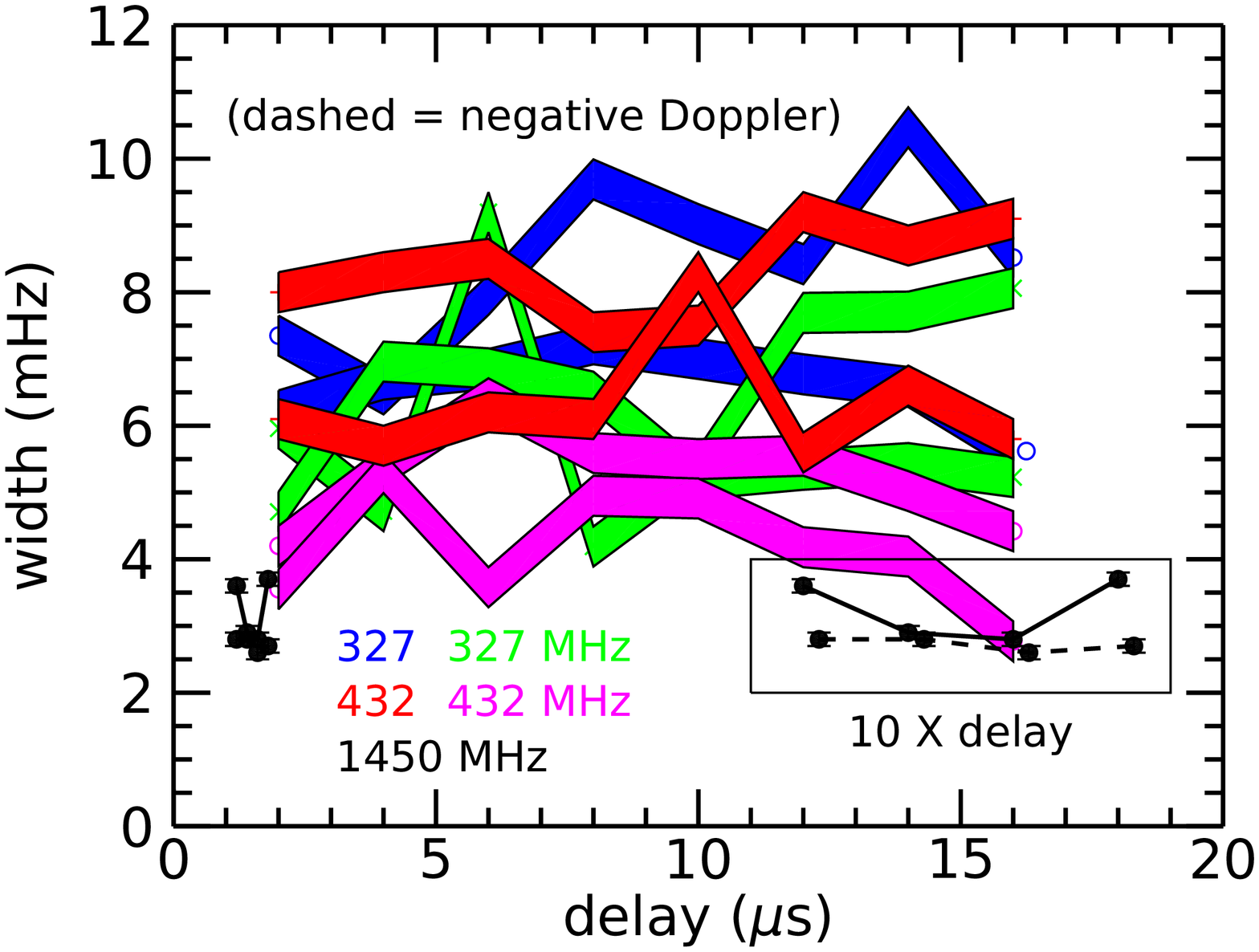}
\includegraphics[width=0.33\textwidth]{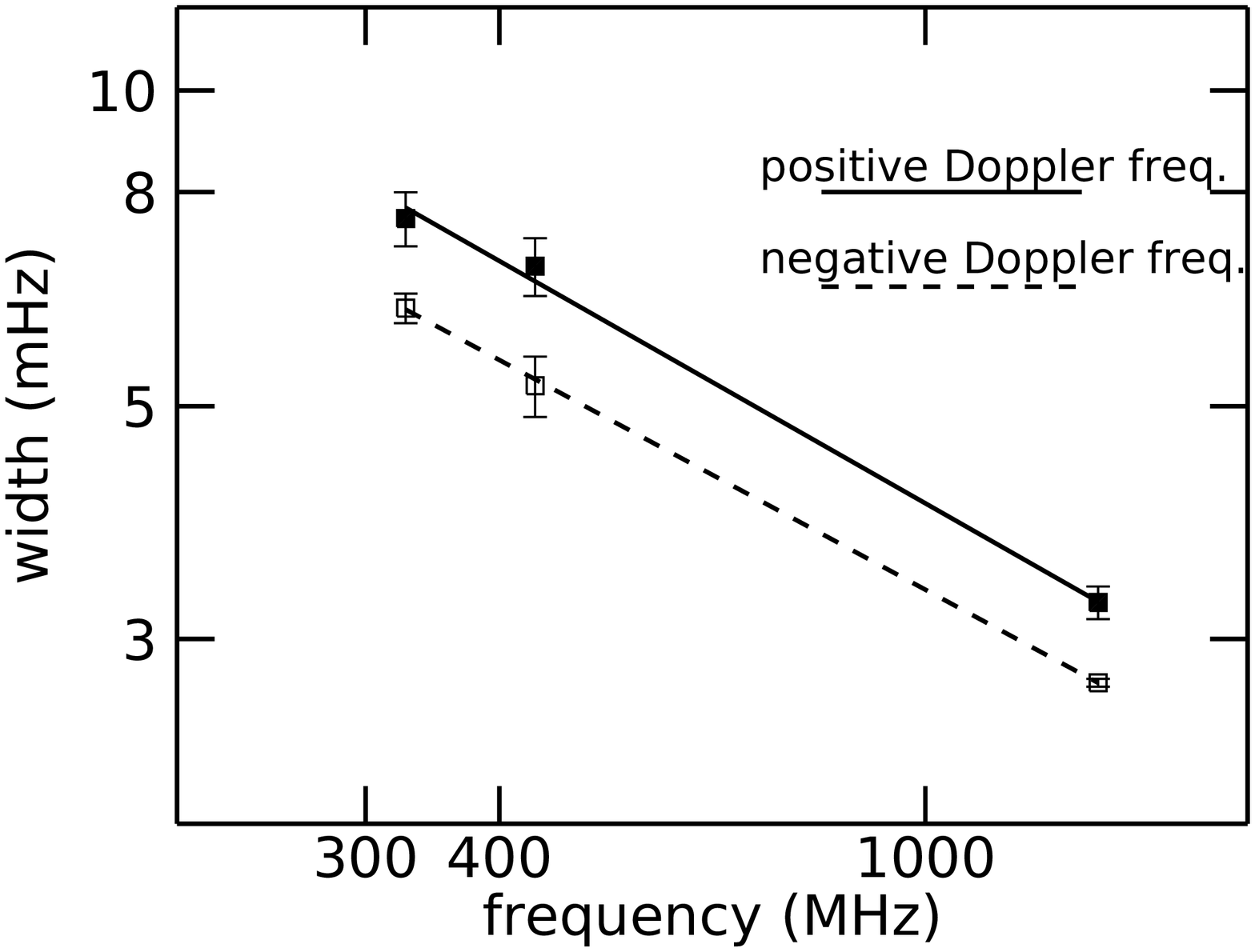}}
\caption{Left: Cross-cut of the scintillation arc (linear power)  at 10 $\mu$sec delay with fitted Gaussians (327~MHz, MJD 57173). Center: Results of crosscut analysis for the five data sets. Plotted are the Gaussian widths to 1/e in fringe (Doppler) frequency vs.\ delay. The shaded areas indicate 1-sigma formal fitting errors for the two lower frequencies. All error bars are shown for the higher frequency along with a boxed inset showing the data with the delay value multiplied by 10. {\bf[Note for the arXiv version: see the .pdf version in the {\tt source} directory or contact DRS.]} Right: Average arc widths as a function of delay (log-log). The data in Figure~2b were averaged in an unweighted fashion to produce an average point at each frequency for the positive and negative Doppler arcs separately. Error bars (1-sigma) are the standard deviation of the mean from the unweighted average. The best fit lines for the positive and negative arcs have slopes $b_+ = 0.58 \pm 0.04$ and $b_- = 0.55 \pm 0.02$, where $\DfD \propto \nu^{-b}$, and the uncertainties are 1-sigma formal errors.
\label{fig:cc_fits}}
\end{figure*}

To measure the scintillation arc thickness in fringe frequency, $\fD$, cross-sectional cuts of the secondary spectrum {\em in linear power} were taken at evenly spaced delay intervals, where each cut averaged the pixel intensity over 20 delay bins. Prior to averaging, the data for each delay bin were appropriately shifted based on the overall arc curvature value $\eta$.
Using a non-linear least squares fit, we determined the Gaussian mean and full-width $\DfD$ to $1/e$ of each peak in intensity.
Figure \ref{fig:cc_fits}a shows a crosscut at fixed delay through the secondary spectrum at 327 MHz and also Gaussian functions that were fitted to them.  

{Because there was no day on which we observed at all three frequencies --- and in order to assess the statistical stability of our analysis --- we analyzed five data sets, as detailed above.
In Figure~2b it is clear that the day-to-day variation in the crosscut width measure is substantial.
Nevertheless, within a given day there is no systematic trend in the values as a function of delay, as implied by Equation~\ref{eq:DeltafD}; this provides support for an essentially 1D brightness distribution.
Since Equation~\ref{eq:DeltafD} indicates that a constant-$\tau$ crosscut maps out $B(\theta)$, we expect that the linear power crosscuts we have performed are mapping out the brightest (core) portion of the image, as we discuss further in \S4.2.

The measured arc thicknesses were averaged to give a single value at each frequency, separately for the positive and negative Doppler portion of the arc.
The averaging was done in an unweighted fashion and included $N = 16$, 16, and 4 points for the averages at 327, 432, and 1450~MHz, respectively.
The error bars in Figure~2c are the standard deviation of the mean of these averages, $\sigma_{\bar{x}} = \sigma/\sqrt{N}$.
The values and formal uncertainties in the logarithmic slopes of these lines
are $b_+ = 0.58 \pm 0.04$ and $b_- = 0.55 \pm 0.02$, where $\DfD \propto \nu^{-b}$.
However, an inspection of Figure~2b indicates that the variation in between closely spaced epochs is not insignificant and is probably caused by ``refractive substructure"  (Johnson and Narayan, 2016) as commented on further in \S5.1.
Hence, we adopt a more conservative error estimate on these slopes, assigning an uncertainty of 2.5 times the larger of these formal uncertainties: $b_+ \approx b_- \approx 0.6 \pm 0.1$.
Thus, we conclude that the logarithmic slope of this quantity is less than 0.8, placing this in substantial tension with theoretical expectations (see further discussion below).


\subsection{Fitting the 1-Dimensional Brightness}
\label{sec:1dfits}

Having estimated  the thickness of the forward arcs in \S\ref{sec:cc}, we now use 
Equations~\ref{eq:S21D}--\ref{eq:1Dtheory} 
to model the entire secondary spectrum, which we fit to the observations.   
This is similar to, but improves upon, the approach used by Trang and Rickett (2007) on earlier Jodrell Bank data for this pulsar.\nocite{tr07}
To define the model, we need the curvature $\eta$ estimated as part of the crosscut fitting; to determine the angles $\theta_{\parallel}$ in the model, we also need $D_{\rm eff}$, which depends on $s$ as defined in section\S\ref{sec:theory}.
In what follows we express most of our results in terms of $\theta_{\rm m}$, which is related to $\theta_{\parallel}$ through Eqn~\ref{eq:theta_m}.

To the 1D model for the secondary spectrum \Stwo we add a constant $S_{\rm 2,noise}$ representing the average noise floor.  Thus our parameters are $B(\theta_{\rm m})$ sampled at uniform intervals $\delta\theta_{\rm m}$, which we set as $\delta\theta_{\rm m} = \sqrt{2c\eta/D_{\rm psr}}\delta \fD$.  The observations typically cover $\pm 40$ mHz in $\fD$ in 256 points, giving about 128 points in $B(\theta_{\rm m})$.   

\begin{figure*}[bth]
\begin{center}
\begin{tabular}{lll}
\includegraphics[trim = 5 0 40 40, clip,angle=0,width=5.5cm]{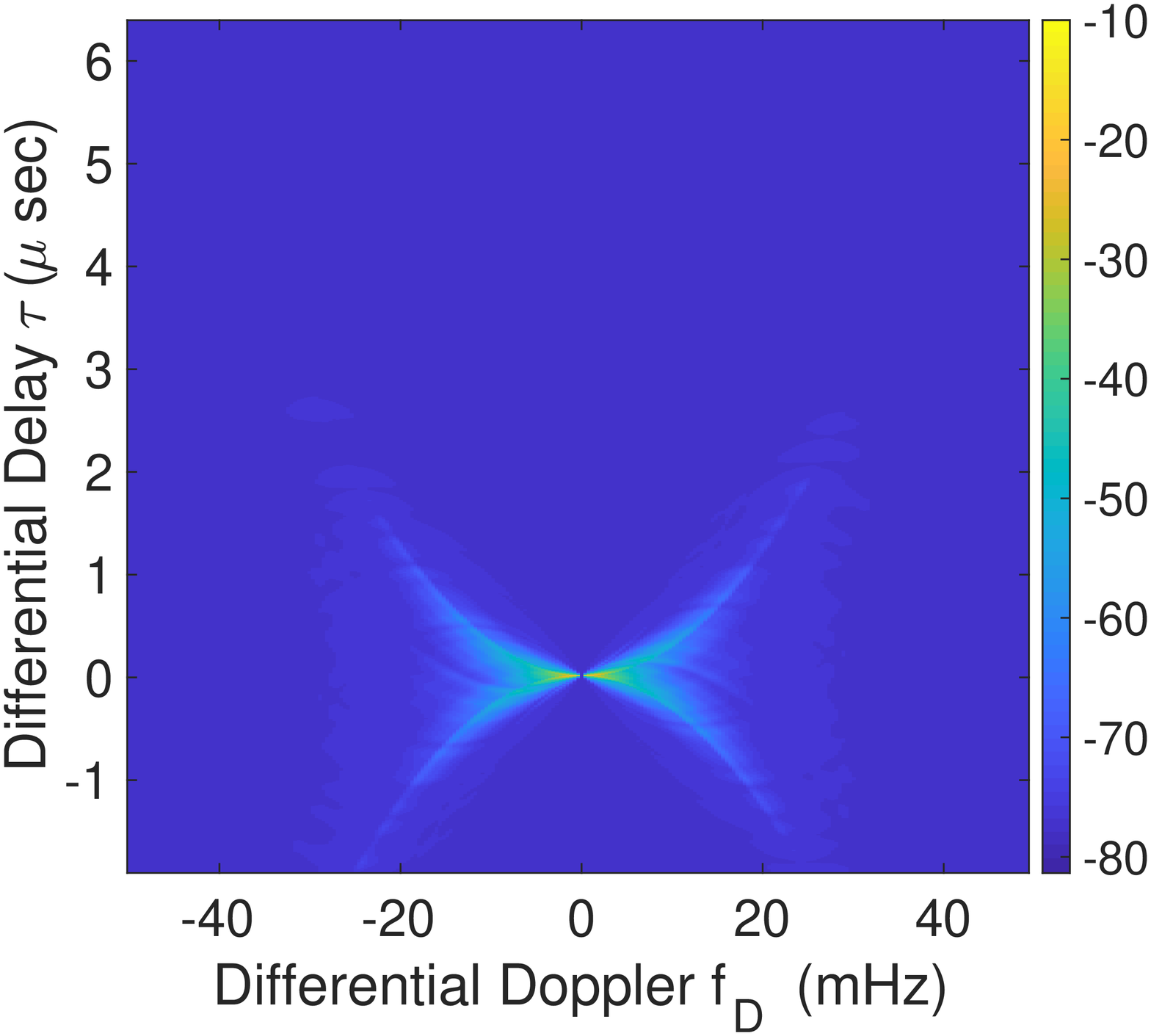} &
\includegraphics[trim = 5 0 40 40, clip,angle=0,width=5.5cm]{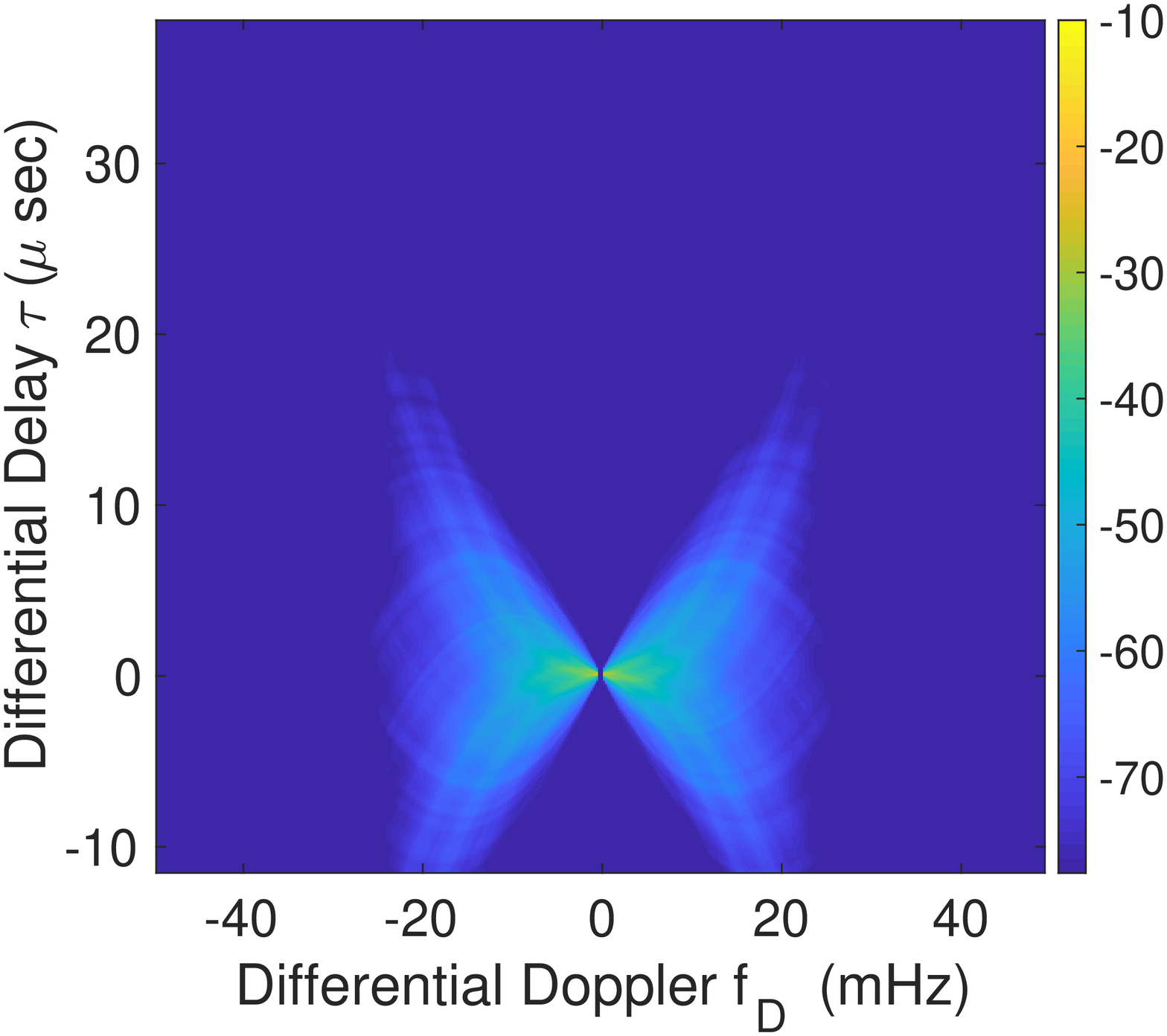}  &
\includegraphics[trim = 5 0 40 40, clip,angle=0,width=5.5cm]{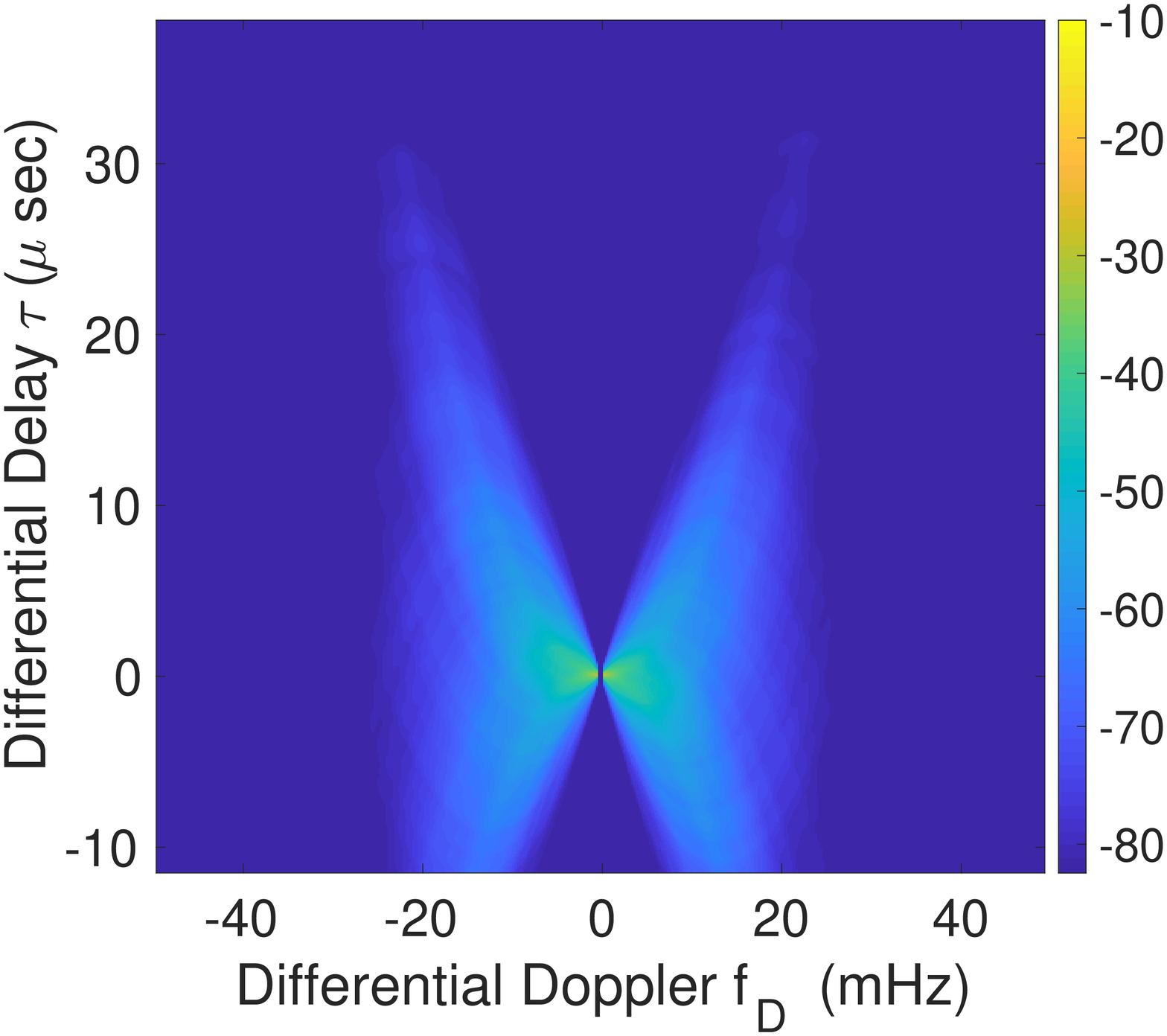}\\
\includegraphics[trim = 5 0 32 40, clip,angle=0,width=5.5cm]{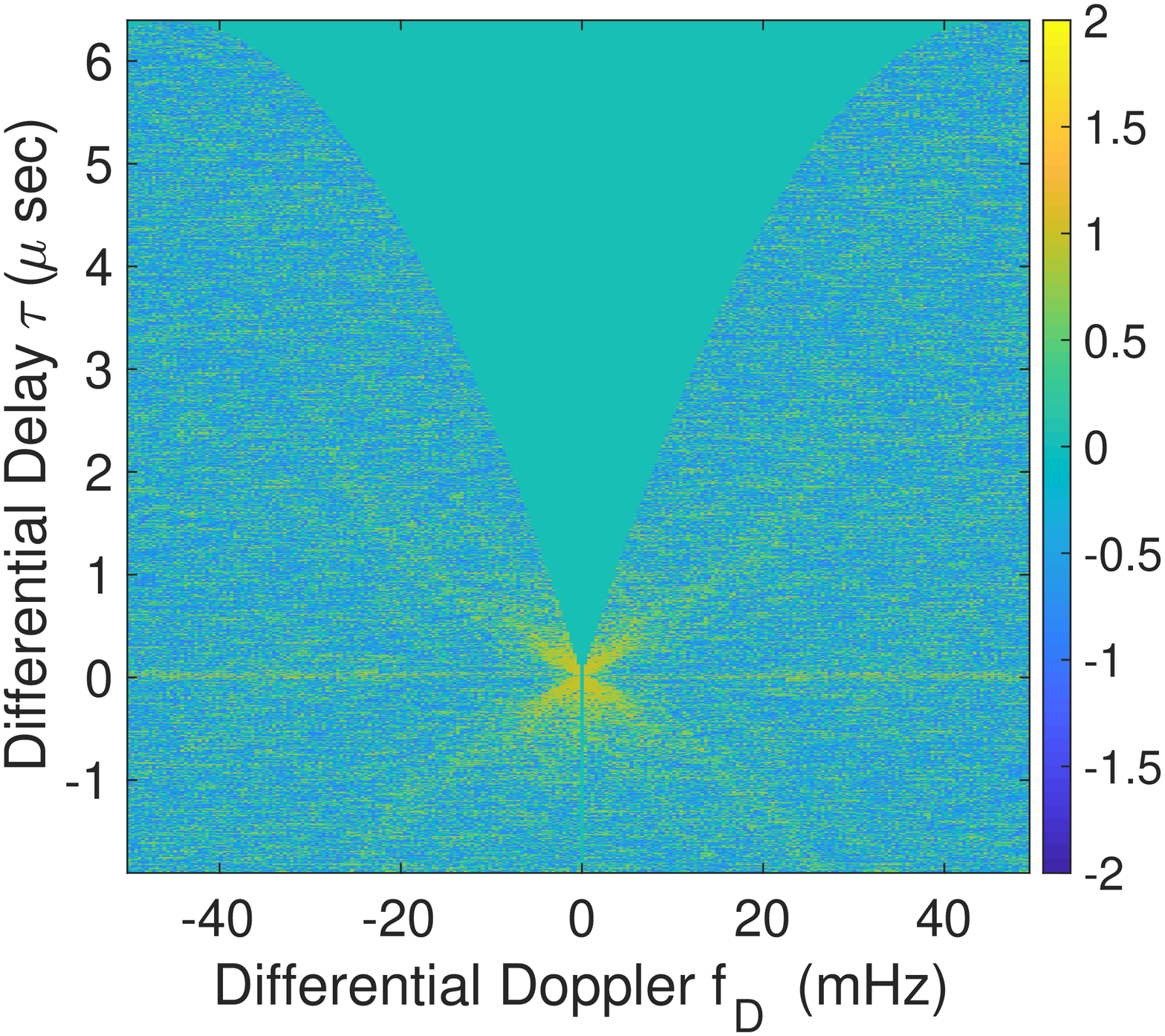} &
\includegraphics[trim = 5 0 32 40, clip,angle=0,width=5.5cm]{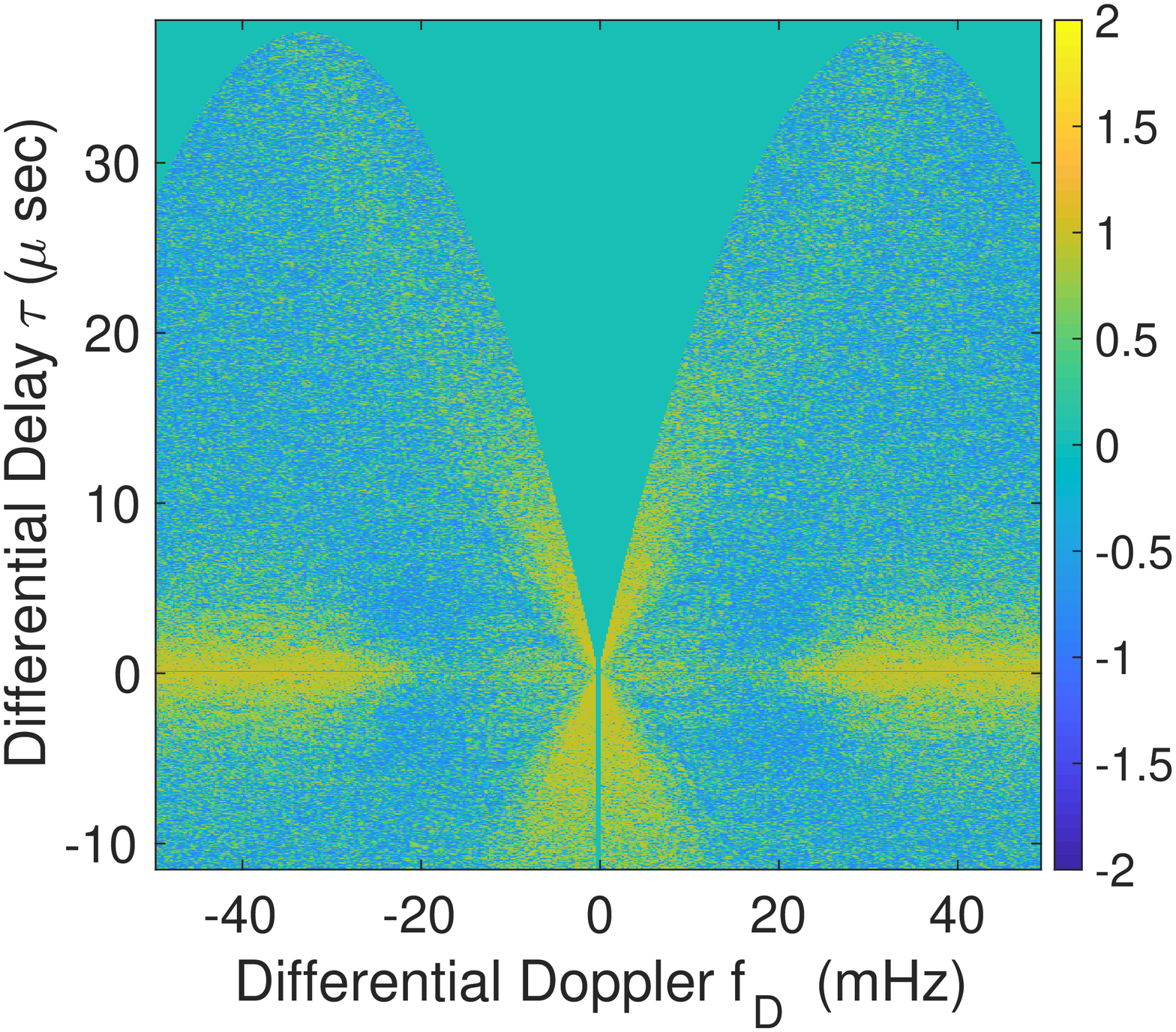}  &
\includegraphics[trim = 5 0 32 40, clip,angle=0,width=5.5cm]{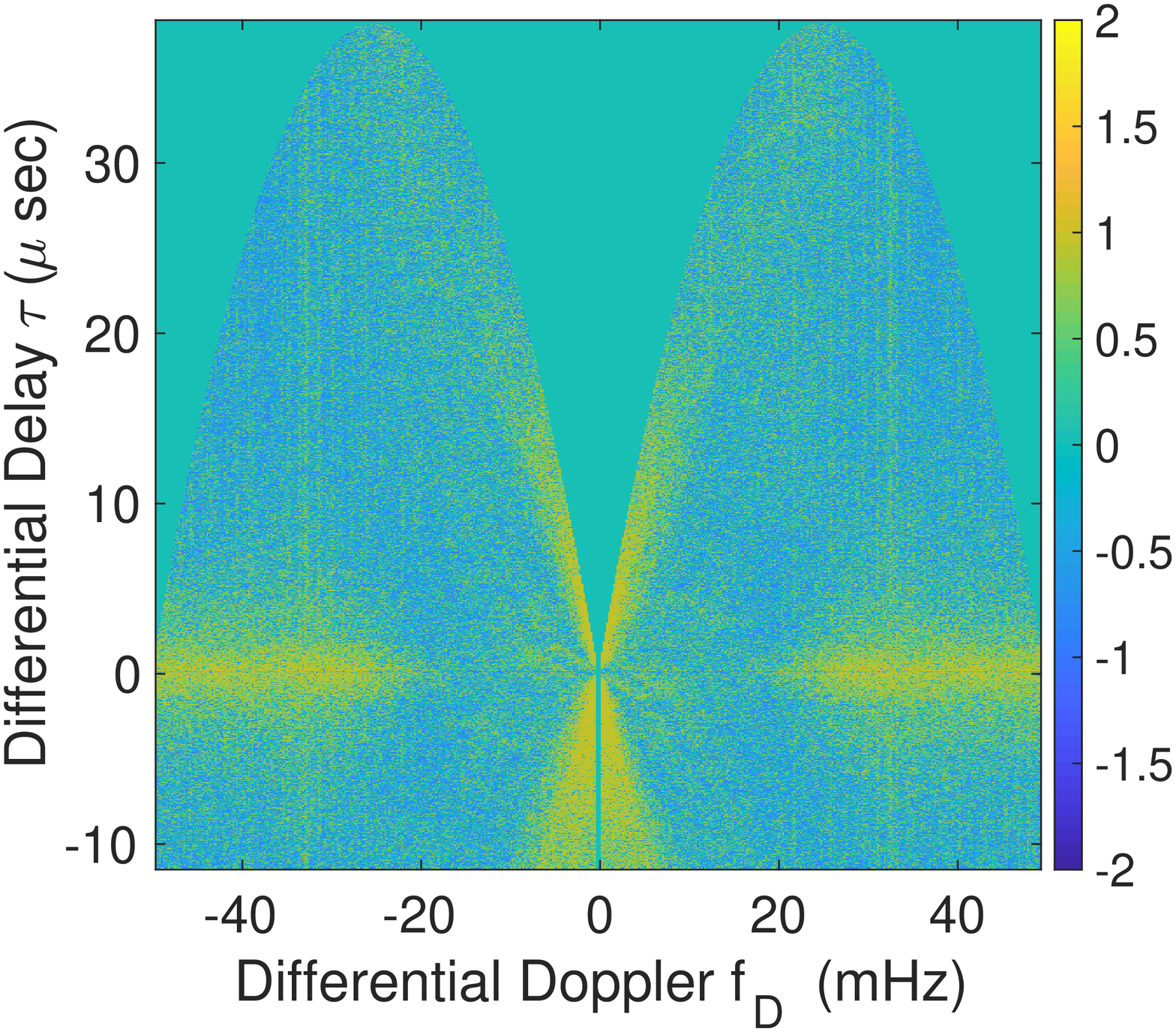}
\end{tabular}
\end{center}
\caption{Upper: models (in dB) fitted to the observed secondary spectra displayed in Figure~\ref{fig:ds}.  Left:  1450 MHz.  Middle:  432 MHz. Right: 327 MHz.  Lower: Residual difference in secondary spectrum between observed and model.   Colorbar scale is (observed - model) normalized by the model.}
\label{fig:1dmodelresA} 
\end{figure*}

The fitting is a least squares minimization of the weighted difference between the observation and the model, 
evaluated over the region where $\tau > 0$ in Figure \ref{fig:ds}.   
The resulting models are in the upper panels of Figure~\ref{fig:1dmodelresA}.  The lower panels show the residual after fitting, which we define as (observed - model) weighted by the reciprocal of the model itself, including the flat noise background.
Thus, the residual is a fraction of the model. This is motivated by the fact that spectral estimates from a single realization of a random process have exponential statistics, in which the standard deviation equals the mean at each point.  
It allows logarithmic scaling in $S_2$  which reveals arcs over such a wide dynamic range.

The models do a reasonable job in representing the data, creating arcs that are thickest at 327 MHz and becoming thinner toward 1450 MHz.  
Note that both model and residuals are set to zero (green) in the widening V-shape centered on the delay axis, which is determined by bounds on $\theta_{\pm}$, set by the maximum values of $\theta$ in the model.   The  yellow V-shaped valley nearby show that the model under-shoots the data near the delay axis, which is the worst discrepancy in the model. These regions would require large brightness at the outer range of $\theta$ in the model.   But higher brightness at large angles would extend the arc out to larger delays than in the observations.  It is clear that the linear 1-D brightness function cannot model the signal in this valley near the origin, for which some significant \it perpendicular \rm  width is needed.   Nevertheless the basic form of the observed arcs are described well, and so it is worth comparing the brightness functions at different frequencies.   Note that there is also some unmodeled signal along the Doppler axis, which is due to residual pulse-pulse modulations that have not been fully corrected.

\begin{figure*}[thb]
\begin{center}
\begin{tabular}{lll}
\includegraphics[trim = 0 5 50 10, clip,angle=0,width=5.5cm]{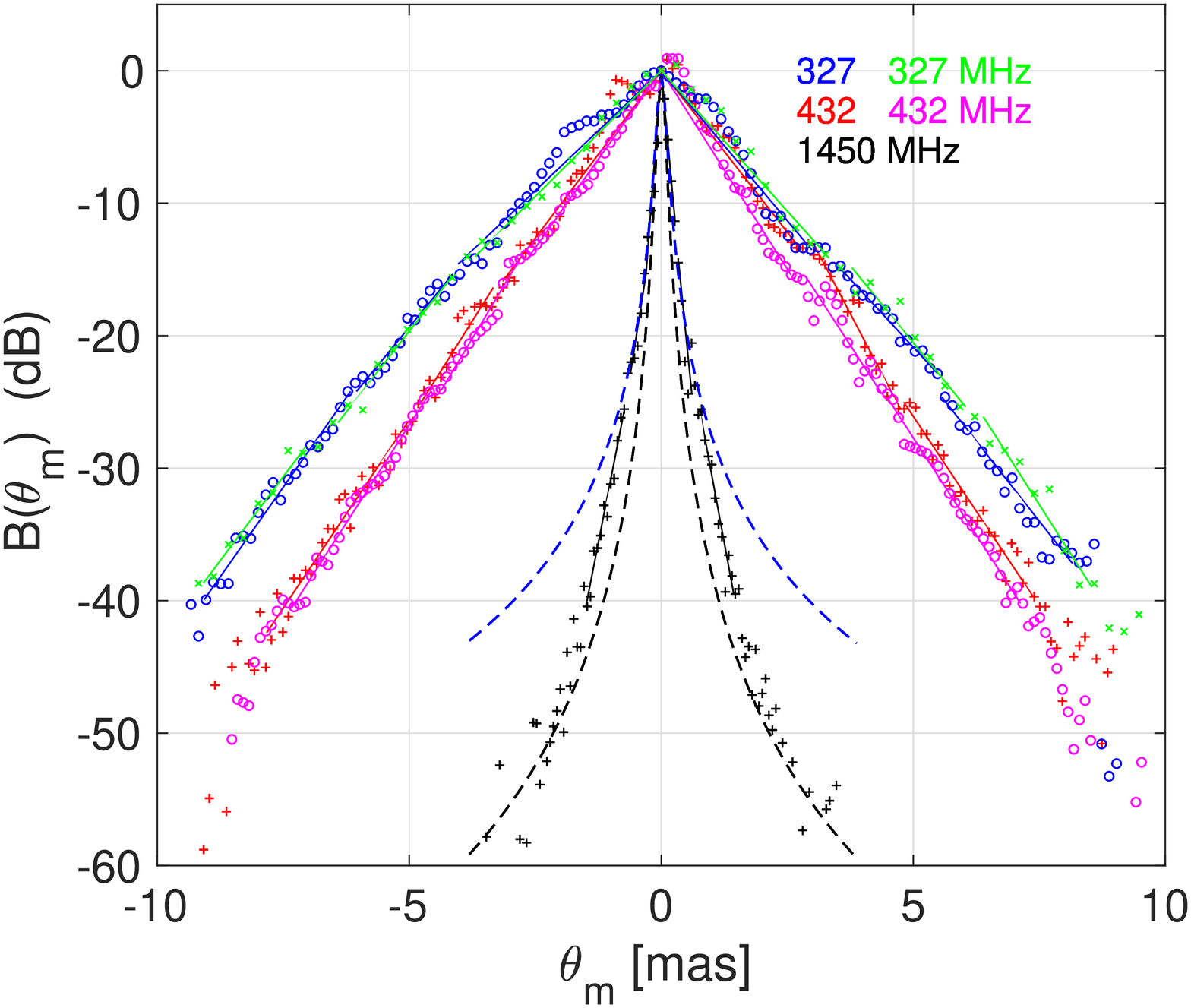}  &
\includegraphics[trim = 5 5 50 10, clip,angle=0,width=5.5cm]{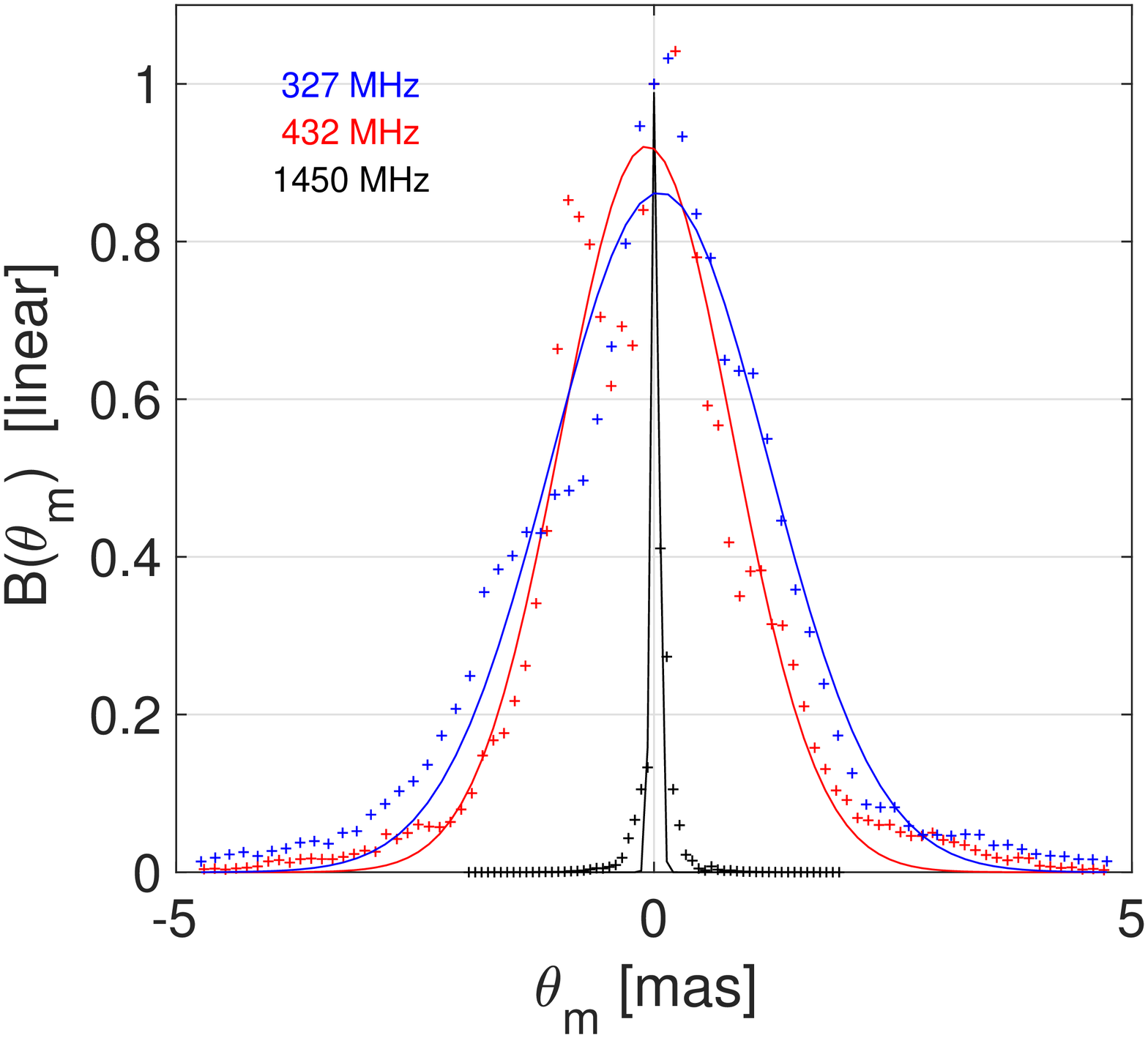} &
\includegraphics[trim = 5 0 40 10, clip,angle=0,angle=0,width=6cm]{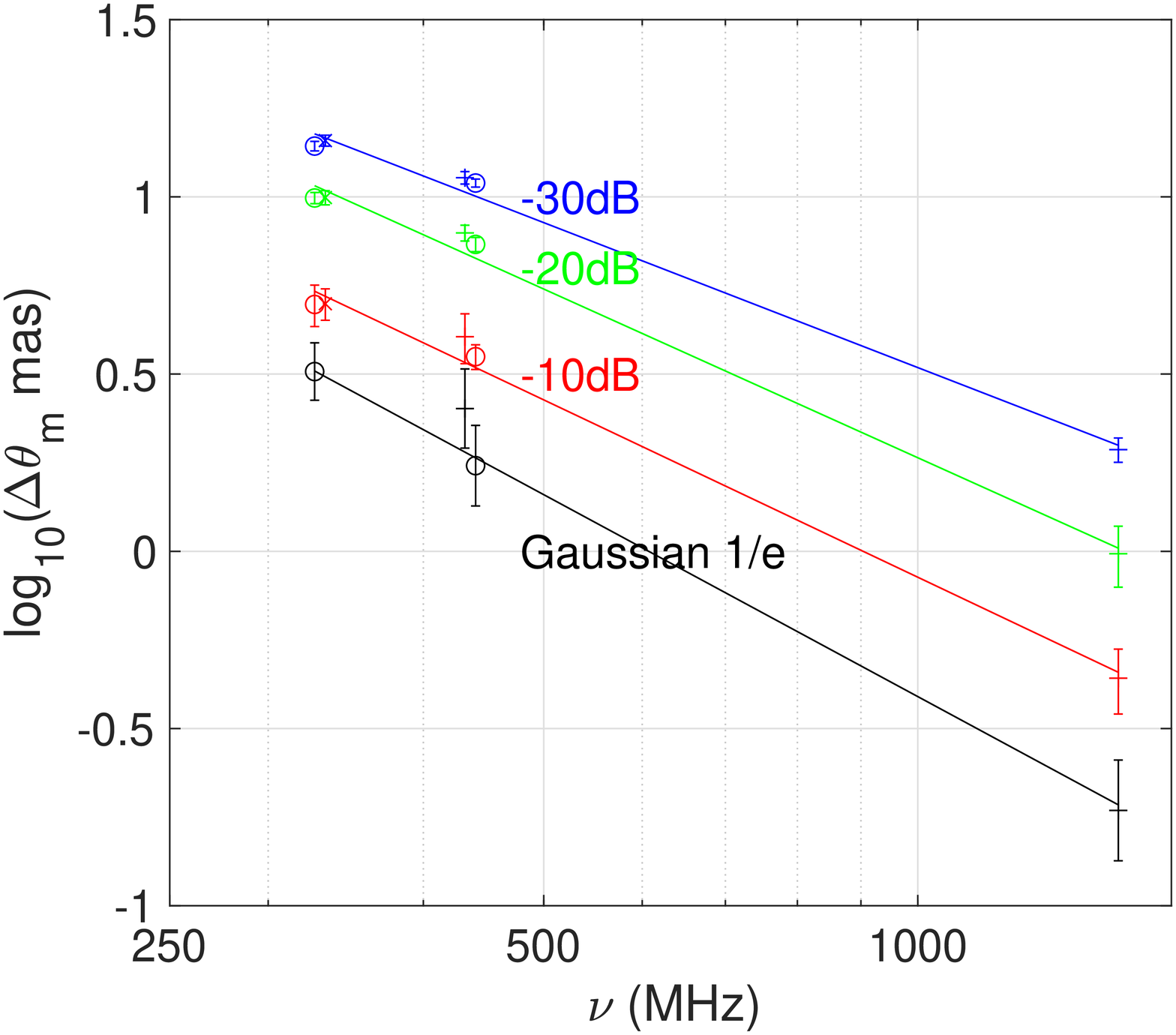} \\
\end{tabular}
\end{center}
\caption{1-D brightness models $B(\theta_m)$  (in dB) fitted to the observed secondary spectrum identified by color at the three frequencies. 
The horizontal axis is a scaled parallel angle $\theta_{m}$  in mas assuming that the scattering takes place at $s_0 = 0.62$.   
Left:  {Symbols identify the MJD of the observations: `$+$' for 57179,  `$\circ$' for 57173 and `x' for MJD 57185}, with straight lines that are fitted to 10~dB ranges centered at -10, -20 \& -30 dB.  
At 1450 MHz, theoretical brightness distributions for isotropic and 1D Kolmogorov density spectrum are over-plotted as dashed lines; these are both constrained to have the same 3 dB width as the fitted brightness, and the black dashed curve is the isotropic case.
Middle: linear plot shows the prominent spike at 1450 MHz; see discussion.  
Right: The full width of $B(\theta)$ was estimated at 3 levels from piece-wise linear fits that covered $\pm 5$~dB in power level centered at -10, -20 \& -30 dB below the peak. The full width of the Gaussian model characterizes the width near the peak.   Data from the same frequency but differing days are offset horizontally.
The straight-line fits at the three levels gave slopes that are listed in Table \ref{tab:bright_widths}.   See text for discussion.
}
\label{fig:Bmodel} 
\end{figure*}

\section{Discussion}
\label{sec:DiscConc}

Our secondary spectra observed from B1133+16 show clear forward arcs whose thicknesses decrease strongly with frequency.   We have characterized this by fitting Gaussian functions to crosscuts through $S_2$ at fixed delay.   The widths $\Delta$$\fD$ do not depend {systematically} on the  delay chosen for the crosscut.  This is readily explained in a 1D brightness model, summarized in Equation~(\ref{eq:DeltafD}), where the scaling constant from $\fD$ to angle is independent of delay.  
Our final assessment of the crosscut frequency dependence is $b \approx 0.6 \pm 0.1$, and we conclude that $b < 0.8$.
Using Equation~(\ref{eq:DeltafD}), {this corresponds to} the characteristic width $\Delta \theta$ of $B(\theta_{\parallel})$ scaling as  $\propto \nu^{-a}$, where the exponent $a =b+1 < 1.8$.
This is {\em significantly} shallower than both $a=2$ expected for plasma refraction and $a=2.2$ for Kolmogorov turbulence in a plasma layer (screen).  

In \S\ref{sec:1dfits} we found satisfactory fits to $S_2$ from 1D models of the scattered brightness distribution at all 3 frequencies, which are shown in the left panel of Figure~\ref{fig:Bmodel}.  {We conclude that the brightness distributions must be highly anisotropic and plot them on a logarithmic scale against $\theta_m$ along the major axis}. 
Evidently  $B(\theta)$ widens considerably with decreasing frequency.   However, one can also see that the shape of  $B(\theta)$ changes somewhat with frequency.  On the log/linear scale the brightness function at 1450 MHz flares out somewhat at low levels, while at 327~MHz it falls more steeply at low levels. 
{We note, however, that the 1450 MHz brightness does not flare out at low levels as much as either the isotropic or 1D Kolmogorov models which are overplotted.}   

{In order to best quantify the shape} of $B(\theta)$ we estimated the width at three levels -10,-20,-30~dB down from the peak at 0~dB.  
{To partially mitigate} the noise-like variations in the estimated brightness functions, we fitted straight lines to segments that span approximately $\pm 5$dB around each level on both the rising and falling sides of the profile.  
The full widths between where the fits crossed each level are plotted on log scales against frequency in Figure \ref{fig:Bmodel} and listed in Table \ref{tab:bright_widths}.  

We also plot the brightness profiles on a linear scale in the middle panel of Figure~\ref{fig:Bmodel}, which reveals their shapes near their peaks.  We characterized the widths of the peaks from the fitted Gaussian functions (on linear scale) as shown by the lines.   
At 1450 MHz the peak is very much narrower than at 432 \& 327 MHz, whose widths are nearly equal.  The results are listed in Table \ref{tab:bright_widths}; notice that the width at 1450 is {19 times} narrower than at 432 MHz.  This spike at 1450 MHz is reminiscent of an unscattered component that might be present when the scintillation is weak; see C+06 for a discussion of arcs in weak scintillation.  The spike gives rise to the narrow ``spine'' along the forward parabola in the left panels of Figure \ref{fig:ds} \& \ref{fig:1dmodelresA}. In the table we also list the overall modulation index ($m=$ rms/mean) derived from each dynamic spectrum.  While at the lower frequencies the scintillations are strong ($m > 1$), the scintillations at 1450 MHz are near the transition to weak ($m = 0.56 < 1$) consistent with an unscattered component in the brightness.
In Figure \ref{fig:Bmodel} the width drops more steeply between 432 and 1450 MHz than between 327 and 432 MHz, as is also seen for the crosscut widths in Figure~\ref{fig:cc_fits}.  We conclude that the steeper scaling from 432 to 1450 MHz is due to the emergence of an unscattered spike component at 1450.  This makes the estimates of $a$ in the table very uncertain.

\begin{table*}[hbt]
\centering
\begin{tabular}{c c c c c c}
Frequency & Modulation  &  Width & Width & Width & Width\\     
{} & index &  Gaussian  &  -10 dB &  -20 dB &  -30 dB \\
(MHz) &   &mas & mas & mas & mas \\
\hline
1450 & 0.56 & 0.18 &  0.44 &  0.98 & 1.9 \\
432 & 1.11  & 2.5 &  4.0 & 7.9 & 11.3\\
327 & 1.10  & 3.2 & 5.0 & 10.0  & 14.0 \\
exponent $a$   &  -  &   1.9$\pm$ 0.2  & 1.7 $\pm$ 0.1  & 1.6$\pm$ 0.1 & 1.4$\pm$ 0.1 \\

\end{tabular}
\caption{Angular widths from Gaussian fits and {width estimates} at 4 different levels in the brightness profiles from Figure \ref{fig:Bmodel}.  Bottom row is the fitted scaling exponent $a$ from the 4 listed widths. }
\label{tab:bright_widths}
\end{table*}

The width results in Table \ref{tab:bright_widths} show that the frequency scaling exponents at -10 and -20 dB are consistent with the value $a=1.7$ found from the crosscuts.  However, the flatter frequency scaling at the lowest level ($a = 1.4$ ) is a separate result, which we discuss in section \ref{sec:freqscal}.   The changes in $a$ estimated at different levels reflect the different shapes visible in Figure \ref{fig:Bmodel}.   
The major observational result from our paper is this frequency dependence in the scattered angular brightness $B(\theta)$.
That our estimates of $a$ are significantly less than both $a=2$ expected for plasma refraction and $a=2.2$ for Kolmogorov turbulence in a plasma layer suggests a comparison with the many recent reports of flatter than expected scaling

Our lower than expected scaling exponents for the width of the angular scattering can be compared to the exponents reported by a number of authors (L\"ohmer et al.\ 2001; L\"ohmer et al.\ 2004; Bhat et al. 2004, Geyer \& Karastergiou 2016; Krishnakumar et al.\ 2017; Geyer et al.\ 2017 and references therein) who have measured the frequency scaling law for the diffractive interstellar pulse broadening time ($\tau_{d}$) in many pulsars.  
\nocite{lkm+01,lmg+04,gk16,kjm17,gkk+17}
Since the scatter broadening time must vary as the square of the angular broadening the scaling exponent is expected to be $\tau_{d} \propto \nu^{-2a}$.  For most pulsars the exponent $2a$ is found to be slightly but significantly less than the canonical values of $4.0$ for plasma refraction or $4.4$ for Kolmogorov plasma turbulence.

\subsection{Diffractive de-correlation bandwidth}

In the upper left panel of Figure 1 the dynamic spectrum at 1450 MHz exhibits large islands of intensity (about 20 MHz by 2 minutes), which are modulated by much finer criss-cross structure. In a traditional auto-correlation analysis of these data the diffractive de-correlaton bandwidth would be estimated as the halfwidth frequency offset $\nu_d$ needed to de-correlate the scintillations by 50\%, with a diffractive time scale estimated from the width in the time domain.  We computed the autocorrelation function (ACF) from the Fourier transform of $S_2(\tau,\fD)$ and found that the decay along the frequency and time axes both exhibited a two scale structure. In addition there are linear features with differing slopes on the two scales, which correspond to the criss-crossing slopes evident in the dynamic spectrum.  Similar results were obtained for all three frequencies.  However, the rich detail revealed by the parabolic arcs in the secondary spectrum is absent in the ACF.   While the theory of strong ISS predicts $\nu_d \propto \nu^{2a}$, the two-scale appearance of our acfs make $\nu_d$ too subjective  to be useful for determining the scaling index from $\nu_d$.  

Even with this uncertainty it is evident that $\nu_d \ll \nu$ at all three frequencies.   Narrow fractional bandwidths are normally the hallmark of strong diffractive scintillation, which raises a conundrum at 1450 MHz where the observed scintillation index $m < 1$.  The explanation lies in the high degree of anisotropy found in section \ref{sec:1dfits}.  For isotropic scattering screens a diffractive scale $r_{\rm d}$ is defined where the structure function of screen phase equals one, i.e.~it is the scale for an rms phase difference of 1 radian.  This gives the typical diffractive scattering angle as $\theta_d = \lambda/( 2\pi r_d)$.  When $ r_{\rm F} \gg r_d$ strong scintillation is caused by the mutual interference between the scattered waves, where $r_{\rm F}$ is the Fresnel scale (e.g.~Narayan, 1992).  The scattered waves have differing delays $\tau_d \sim D_{\rm eff} \theta_d^2/(2c)$ which cause the deep modulations in the intensity spectrum over fequency differences $\nu_d \sim 1/(2\pi\tau_d)$ and $\nu/\nu_d \approx (r_{\rm F}/r_d)^2 \gg 1$.  

However, when the scattering is highly anisotropic the diffractive angles along the major and minor axes differ by a large axial ratio.  Under these circumstances it is the greater width along the major axis of angular scattering that sets $\nu_d$ which can be much finer than $\nu$, even though the intensity fluctuations are not fully developed along the minor axis and overall $m$ can be less than 1.

Before discussing the scaling results in a wider context, we consider the possible influence of time variations in the scattering medium, which might enter because the 5 data sets span a 15~day interval.
We  fitted the 1D model to all the observations reported here.  
Thus in the left panel of Figure \ref{fig:Bmodel} together with MJD 57179 we have overplotted two dates at 327 and two dates at 432 MHz (57173 being common to both).  One can see the somewhat wider profiles at the lower frequency and also the effect of time variations at both frequencies.  

The time variations are chiefly visible as ripples of a few dB on angular scales comparable to the 3dB widths.  The 45-60 minute duration of each observation is definitely shorter than the estimated refractive timescale, and thus each arc observation will be in the ``average image'' mode (see Johnson and Narayan, 2016) as distinct from the ensemble average image.   Thus we suggest that the ripples are examples of refractive substructure (Johnson and Narayan, 2016), which raises a further possibility.   The peaks and valleys in our estimated  brightness are the cause of reverse sub arcs, which are due to the same substructure.   In addition it is likely that most estimations of scatter broadening times $\tau_d$ will also be subject to this type of refractive estimation error and so the fitting of a frequency scaling index may have higher errors than quoted. 
}

\subsection{Comparison with frequency scaling laws from the literature}
\label{sec:freqscal}

{Our shallower than expected frequency scaling for the width of the angular scattering can be compared to the frequency scaling} exponents reported by a number of authors (L\"ohmer et al.\ 2001; L\"ohmer et al.\ 2004; Bhat et al. 2004, Geyer \& Karastergiou 2016; Krishnakumar et al.\ 2017; Geyer et al.\ 2017 and references therein) who have measured the frequency scaling law for the diffractive interstellar pulse broadening time ($\tau_{d}$) in many pulsars.  
\nocite{lkm+01,lmg+04,gk16,kjm17,gkk+17}
Since the scatter broadening time must vary as the square of the angular broadening the scaling exponent is expected to be $\tau_{d} \propto \nu^{-2a}$.  For most pulsars the exponent $2a$ is found to be slightly but significantly less than the canonical values of $4.0$ for plasma refraction or $4.4$ for Kolmogorov plasma turbulence.  

A possible explanation of this discrepancy was proposed by Lazio \& Cordes (2001).
\nocite{cl01}
They examined the effect of scattering in screens of a finite transverse dimension.  The basic idea is that plasma scattering or refraction increases at long wavelengths, causing a broadening in angle $\propto \nu^{-2}$.  
Thus, there is a ``scattering disk,'' which increases in diameter $\propto \nu^{-2}$, at the screen from which the observer receives the waves.   
For screens of finite dimension, when the scattering disk exceeds that dimension in at least one direction, the range of angles received is cut off causing a slower than $\nu^{-2}$ scaling at the lowest frequencies.  This idea could explain why we find a scaling exponent less than 2 and why the outer edges of the brightness distribution falls more steeply at 327 MHz than at 1450.  However, there is no sharp cut-off in our brightness models, and so an explanation would have to invoke transverse variations in the strength of the turbulence (or whatever causes the scattering).  Such transverse variations could also cause asymmetrical increases or decreases in brightness that are time variable. Under such a scenario there would not be a global frequency-scaling law.   

This discussion connects with the work of Geyer et al.\ (2017), who have {analyzed} the frequency-scaling law of $\tau_d$ for a large sample of mid- to high-DM pulsars.  
In most cases they find flatter than expected scaling exponents ($<4$).  However, they also conclude that much of the discrepancy may be due to the improper use of pulse broadening functions that assume isotropic scattering when the underlying plasma may be highly anisotropic.   
This is connected to our finding of very anisotropic scattering in B1133+16 and the well-documented case of extreme anisotropy of scattering for B0834+06 (Brisken et al.\ 2010).  
In future work we will compare the pulse broadening function derived from the 1D brightness models in B1133+16 with the various theoretical models that have been used in estimating $\tau_d$.   Geyer et al.\ (2017) also consider the effect of truncation by finite scattering screens, as discussed above.  It appears that there is population of relatively small (AU scale) enhancements in plasma density (and turbulence) that become increasingly common in the inner Galaxy.   For their heavily scattered pulsars one must study the cumulative effect of many screens of finite extent.

{By contrast to these statistical studies of time-domain scattering}, we present a detailed {frequency-domain} analysis for one pulsar and note some particular features of the line of sight towards B1133+16.   It is at very high ($69^{\circ}$) Galactic latitude and seen through the Local Bubble (see Bhat, Gupta \& Rao 1998;  Yao, Manchester \& Wang, 2016).   
\nocite{bgr98,ymw17}
From these two models for the Local Bubble and its boundary, the plasma density along the path toward B1133+16 is very low until the edge of the Local Bubble at about 170 and 250 pc for the two models, respectively.   From our measurements of the arc curvature $\eta$, we can constrain the distance to the scattering screen that causes the arcs using Equation~(\ref{eq:1Dtheory}), where the effective velocity and distance are given above (Brisken et al., 2002).  Although the angle $\psi$ between the 1D scattered axis and the effective velocity is not known, the measured $\eta$ gives an upper limit to the fractional screen distance $s\leq 0.62$.  
This value agrees closely with the ``b'' arc reported by Putney and Stinebring (2006; arc 2 of Stinebring 2006) for B11133+16 over many years.   
In summary $s\leq 0.62$ implies that the distance from the Earth to the scattering screen $D_{\rm s} \geq 136$~pc, and so is consistent with scattering from a concentration of plasma at the edge of the Local Bubble at either 170 or 250 pc.

While the 1D brightness model succeeds in fitting many features of the secondary spectrum from B1133+16, it fails in the V-shaped region near the delay axis, as shown by the residual plots in Figure~\ref{fig:1dmodelresA}.  The observed $S_2$ is filled in at a low level in this region. The region is sensitive to any extent of the brightness function perpendicular to the 1D axis, {which might have the effect of biasing the 1D fit at low levels as the model tries to reduce the differences  (yellow) from the observations.   Thus the frequency dependence of the shape at low levels must be regarded as a preliminary result.}
In future work we will pursue ways to model the brightness in 2D and the need to go beyond the concept of a single axial ratio.  There is considerable potential for a 2D modeling capability from pulsar recordings at a single antenna.

Lastly, we discuss implications regarding the physical structures that cause the arcs.  We measure a widening of arc thickness and of the brightness profile with decreasing frequency, which should  be contrasted with the frequency independence of the reverse arclets in B0834+06, first reported by Hill et al. (2005) and also seen by Brisken et al.\ (2010).  We assume that the same basic phenomenon is responsible for both the forward arcs and reverse arclets in both pulsars and consider how to reconcile their differing frequency dependencies.  Two different types of plasma structures have been proposed.  

Pen \& Levin (2014) and Simard \& Pen (2017) suggest refraction at ripples in a sheet of plasma, such as a current sheet (or even a thin region of reduced plasma density).  Near-grazing incidence on a current sheet is proposed, in particular.   This model identifies the origin of each arclet at a specific ripple, which would cause refracted rays whose observed angle of arrival would be independent of frequency.  No detailed modeling has yet been published for an overall brightness distribution; but, since it relies on plasma refraction, which is strongly frequency dependent, we expect the individual ripples to be modulated by an envelope whose width would also be strongly frequency dependent. 

{There are two papers that propose models based on extremely anisotropic Kolmogorov turbulence.   The first by 
Brisken et al.\ (2010) emphasizes} their detection of a very elongated scattered brightness distribution from B0834+06.  They interpret it as scattering from a localized region of plasma with anisotropic Kolmogorov turbulence.  Their derived brightness has many narrow peaks that map to individual arclets and implies clumping on an extremely fine scale $(< 0.1$)~AU.  The location of these clumps would cause arclets whose angular position is frequency independent.  One model mentioned is a flux rope (Zheng and Hu, 2018), as occasionally detected in the solar wind, which would have to be more than 10 AU in length with localized ``knots'' that cause the arclets.
In the context of  turbulent plasma models for the scintillation, there is another possible explanation for a high angle cut-off in the brightness distribution.  The turbulent energy cascade may be cut off by a dissipation mechanism at a small ``inner'' scale, which will in turn cut off the high angle scattering (Spangler \& Gwinn, 1990; Rickett et al. 2009).
\nocite{sg90,rjtr09}

The second turbulence proposal is by Tuntsov et al.\ (2013) who plot the 1D brightness distribution from the analysis by Walker et al.\ (2008) of arclets observed from pulsar B0834+06;  their method is analogous to our 1D model fitting.)   They then simulated brightness distributions from a set of models for the spectrum of the plasma inhomogeneities.  They model the spectrum of the column depth-integrated phase of an incident radiowave as a 1D power-law versus transverse wavenumber. They explore power-law indices ranging from -2 to -4.5 and find the best qualitative agreement for indices between -2.5 and -3, in the context where a Kolmogorov spectrum has an index -2.67.  Thus they suggest  a 1D Kolmogorov spectrum as a viable model for the column depth-integrated phase, which could arise from 2D (sheet-like) Kolmogorov turbulence viewed edge-on.  {However the predicted frequency scaling index for the 1D Kolmogorov spectrum is the same ($a=2.2$) as for the isotropic model, and so it is not consistent with the results reported here.}

Both the sheet and turbulence models could yield a broad envelope of scattered brightness that would be frequency dependent, which modulates the fine sub-substructure observed at frequency-independent angles.  Future theoretical work should include quantitative analysis for the frequency dependence that can be compared with the results reported here and used to distinguish between the models.

\nocite{bcc+04,zh18,tbw13,jn16}

\nocite{hs08,pmdb14,s07,s07b}

\vspace{0.3in}
We thank the dedicated staff of the Arecibo Observatory for help with the observations.
During the course of the observations (2015) the following was appropriate: ``The Arecibo Observatory is a facility of the National Science Foundation (NSF) operated by SRI International in alliance with the Universities Space Research Association (USRA) and UMET under a cooperative agreement.'' 
This work was supported by NSF grant 1313120 to Oberlin College and by NSF Physics Frontiers Center award 1430284 to NANOGrav.
Analyses were conducted on SCIURus, the Oberlin College HPC cluster (NSF MRI 1427949).
This research has made use of NASA's Astrophysics Data System.
We thank S.\ McSweeney (see Bhat et al.\ 2016) for the development of  \texttt{parabfit} used in our analysis.
The paper was improved by helpful comments from an anonymous referee.


\end{document}